\documentclass[lettersize,journal]{IEEEtran}
\usepackage{amsmath,amsfonts}
\usepackage{array}
\usepackage{tikz}
\usepackage{amsmath,mathtools}
\usepackage{amssymb}
\usepackage{tikz}
\usetikzlibrary{arrows.meta,chains,shadows,backgrounds,fit}

\usepackage{enumitem}
\usepackage{booktabs}
\usepackage{graphicx}
\usepackage{amsmath}
\usepackage{pgfplots}
\usepackage{algorithm}
\usepackage{subfig}
\usetikzlibrary{shadows, positioning, arrows.meta, calc, chains, fit, backgrounds, fadings}
\usepackage{booktabs}
 \usepackage{tabularx}
 \usepackage{array}

\usepackage{algpseudocode}
\usepackage{boxedminipage}
\usepackage{fancyhdr}
\usetikzlibrary{arrows.meta, positioning, calc, shadows, backgrounds, fit}
\usepackage{multirow}
\usetikzlibrary{arrows.meta, chains, shapes.symbols, shadows}

\usepackage{mathptmx}
\usepackage{array}
\usepackage{textcomp}
\usetikzlibrary{positioning}
\usetikzlibrary{arrows.meta}
\usetikzlibrary{calc}
\usetikzlibrary{shapes.geometric}
\usetikzlibrary{fit}
\usepackage{stfloats}
\usepackage{csquotes} 
\usepackage{url}
\usepackage{verbatim}
\usepackage{cite}
\usepackage{eqnarray}
\usepackage{stfloats}
\usepackage{url}
\usepackage{verbatim}
\usepackage{graphicx}
\def\BibTeX{{\rm B\kern-.05em{\sc i\kern-.025em b}\kern-.08em
    T\kern-.1667em\lower.7ex\hbox{E}\kern-.125emX}}
\usepackage{balance}
\usepackage{pifont}

\begin{document}
\title{Kraken*: Architecting Generative, Semantic, and Goal-Oriented Network Management for 6G Wireless Systems}

\author{
    Ian F. Akyildiz, \textit{Life Fellow, IEEE} and Tuğçe Bilen, \textit{Member, IEEE}
    \thanks{Ian F. Akyildiz is with Truva Inc., Alpharetta, GA 30022, USA (e-mail: ian@truvainc.com).}%
    \thanks{Tuğçe Bilen is with the Department of Artificial Intelligence and Data Engineering, Istanbul Technical University, Istanbul, Turkey (e-mail: bilent@itu.edu.tr).}%
    \thanks{*KRAKEN (Knowledge-centric, Reasoning, And goal-oriented Knowledge NetworK): Inspired by the mythological sea creature, Kraken symbolizes a vast, interconnected intelligence capable of multi-layered reach and collective coordination across the 6G landscape.}
    \thanks{A condensed version of this work, focusing on the architectural vision and case studies, has been submitted to IEEE Communications Magazine under the title "The Network That Thinks: Kraken* and the Dawn of Cognitive 6G."}
}


\maketitle
\begin{abstract}
Sixth-generation (6G) wireless networks are expected to support autonomous, immersive, and mission-critical services that require not only extreme data rates and ultra-low latency but also adaptive reasoning, cross-domain coordination, and objective-driven control across heterogeneous entities over distributed edge–cloud infrastructures. Current AI-enabled network management remains predominantly data-centric, relying on discriminative models that optimize intermediate quality-of-service metrics under quasi-stationary assumptions without explicitly reasoning about long-term service objectives. This article advocates a transition from bit-centric communication toward knowledge-centric coordination in 6G systems. We examine how semantic communication embeds task relevance and contextual importance into transmission, enabling prioritization of meaning over raw data delivery. We further position generative artificial intelligence as a structured reasoning mechanism that synthesizes adaptive policies from knowledge representations aligned with dynamic service intents. Network optimization is therefore reformulated around goal-oriented performance metrics capturing application-level outcomes and intent satisfaction rather than solely protocol-level indicators. To operationalize these principles, we introduce {Kraken}, a unified multi-agent architecture comprising a Knowledge Plane, a distributed Agent Plane, and a semantic-aware Infrastructure Plane. Kraken integrates semantic communication, generative reasoning, and goal-oriented optimization over a distributed knowledge substrate, enabling scalable collective intelligence across heterogeneous 6G domains. Beyond a conceptual vision, Kraken defines concrete functional components including semantic-aware protocol mechanisms, Generative Network Agents, and a knowledge-graph-based semantic substrate, while outlining an evolutionary deployment path from current 5G infrastructures toward knowledge-native 6G systems.
\end{abstract}


\thispagestyle{fancy}

\pagestyle{fancy}
\fancyhf{}
\fancyhead[C]{\normalsize Submitted to Proceedings of IEEE, 2026}
\renewcommand{\headrulewidth}{0pt}

\section{Introduction}

\begin{figure}[h]
\centering
\includegraphics[width=\linewidth]{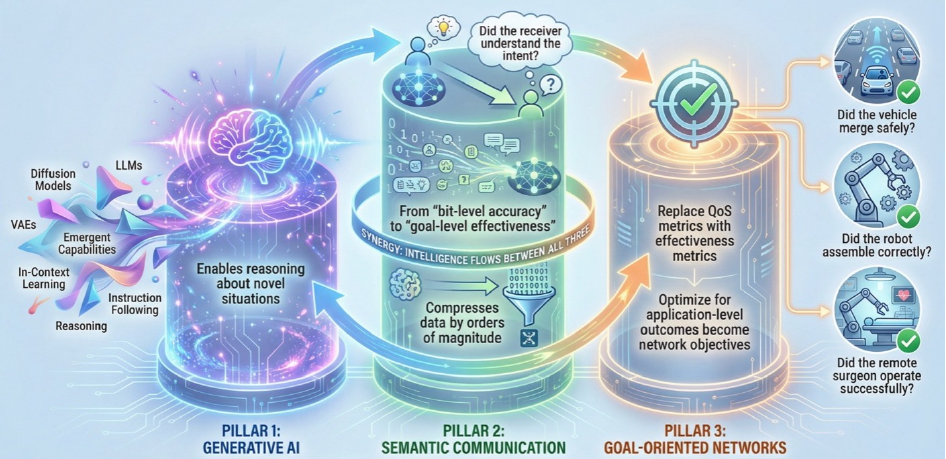}
\caption{The transition from data-centric to knowledge-centric networks relies on three complementary pillars: semantic communication (meaning-aware representation), generative reasoning (world-model-based adaptation), and goal-oriented optimization (intent alignment). Their structured integration enables distributed collective intelligence through coordinated semantic, predictive, and intent-consistent control mechanisms.}
\label{f1}
\end{figure}

Sixth-generation (6G) networks are expected to support tightly coupled communication, computation, sensing, and control across highly dynamic and heterogeneous environments. Emerging applications such as distributed autonomy, immersive XR, collaborative robotics, and large-scale digital twins no longer treat the network as a passive transport substrate \cite{Akyildiz2022_XR,Akyildiz2022_Holographic}. Instead, the network increasingly becomes an active participant in task execution, distributed decision-making, and system-level coordination \cite{BILEN2026111941}. As these systems evolve toward closed-loop cyber–physical architectures, the network must not only deliver data reliably but also contribute to timely interpretation, coordinated adaptation, and goal-consistent control across distributed entities.

Despite this evolution, most existing network intelligence frameworks remain fundamentally data-centric \cite{MartinezJulia2024_AIFramework}. Resource allocation, scheduling, and control decisions are typically optimized with respect to rate, latency, and reliability metrics, implicitly assuming that accurate bit delivery is sufficient for application success. While effective for throughput-oriented services, this abstraction becomes limiting when the utility of transmitted information depends on its contribution to downstream inference, coordination, or control tasks rather than its raw fidelity. In ultra-dense, multi-tenant, and mission-critical environments, identical network-layer metrics may correspond to substantially different task-level outcomes. This mismatch reveals a structural gap between protocol-level optimization and application-level objectives.

Semantic communication has emerged as a promising direction to address this limitation by incorporating meaning-awareness into the transmission process. Rather than treating all bits uniformly, semantic methods prioritize information according to its relevance for downstream tasks, thereby improving robustness and resource efficiency under constrained conditions \cite{10843783}. However, semantic relevance is inherently contextual and must be quantified using task-utility-aware metrics rather than purely bit-level distortion measures. As user intent, environmental state, and service objectives evolve, the semantic mapping itself becomes dynamic. Consequently, semantic representation alone does not guarantee adaptive or goal-consistent behavior in non-stationary environments.

Generative reasoning complements semantic representation by enabling structured adaptation through predictive modeling and model-based policy synthesis \cite{ALBASHRAWI2025100751}. By maintaining internal world models of network dynamics and task interactions, distributed agents can anticipate future states, evaluate alternative coordination strategies, and adapt control policies under uncertainty. Such world-model-based predictive control mechanisms allow network entities to reason beyond instantaneous observations and to proactively regulate system behavior. Nevertheless, predictive intelligence alone does not ensure consistency with operator-defined intents or system-wide objectives. Without explicit representation of service goals and coordination constraints, generative optimization may converge toward locally coherent yet globally misaligned decisions.

In complex 6G ecosystems, intelligence must therefore be both semantic-aware and explicitly goal-aligned. Moreover, given the scale, heterogeneity, and autonomy of future deployments, such intelligence cannot rely solely on centralized orchestration. Distributed agents operating across edge, access, and core domains must coordinate through structured knowledge exchange mechanisms that maintain semantic consistency and intent coherence throughout the system. This coordination requires explicit modeling of goal alignment error, the measurable deviation between local agent actions and global service intents, as a controllable system variable rather than an implicit outcome.

As illustrated in Fig.~\ref{f1}, achieving distributed collective intelligence in 6G requires the structured integration of three complementary capabilities: semantic communication for meaning-aware representation, generative reasoning for predictive and adaptive control, and goal-oriented optimization for intent alignment. Rather than treating these capabilities as isolated enhancements, their joint orchestration establishes a knowledge-centric control paradigm in which communication, inference, and coordination are intrinsically coupled and evaluated against task-level objectives. This paradigm shifts network intelligence from reactive resource management toward proactive, knowledge-driven coordination across distributed network entities.

Motivated by this perspective, we propose \textit{Kraken}, a unified architecture that systematically integrates semantic representation, generative reasoning, and intent-driven coordination within a distributed collective framework. (A high-level overview of the Kraken architecture emphasizing architectural design and representative use cases is presented in \cite{Akyildiz2026KrakenComMag}). The distinctive feature of Kraken is the introduction of an explicit knowledge plane that maintains structured knowledge objects, intent descriptors, contextual state representations, and cross-agent consistency mechanisms across heterogeneous domains. By organizing network intelligence into tightly coupled infrastructure, agent, and knowledge functions, Kraken enables semantic-aware transmission, world-model-based reasoning, and intent-consistent coordination to operate as a coherent system rather than as independent optimizers. Collective intelligence thus emerges from structured semantic exchange, predictive adaptation, and regulated goal alignment across distributed entities. The main contributions of this article are summarized as follows:

\begin{itemize}

\item \textbf{From Data-Centric to Knowledge-Centric Networking:}  
We formalize the transition from bit-level QoS optimization toward knowledge-centric networking, where information is prioritized according to task-utility contribution and system-level objectives rather than raw fidelity.

\item \textbf{World-Model-Based Generative Intelligence for Network Control:}  
We introduce predictive world models and generative policy synthesis mechanisms into network management, enabling adaptive control under heterogeneous and non-stationary operating conditions.

\item \textbf{Intent-Driven Distributed Coordination with Measurable Goal Alignment:}  
We establish goal alignment error as an explicit coordination metric and develop mechanisms that regulate cross-agent consistency through structured knowledge exchange.

\item \textbf{Kraken Architecture for 6G Collective Intelligence:}  
We present a unified multi-plane architecture that integrates semantic communication, generative reasoning, and intent-consistent coordination via knowledge-plane-spanning infrastructure, agents, and knowledge domains.

\item \textbf{Practicality Analysis and Evolutionary Deployment Path:}  
We analyze scalability–complexity trade-offs of knowledge-centric networking and outline an evolutionary transition path from current 5G systems toward Kraken-enabled 6G deployments.

\end{itemize}

To ensure conceptual clarity and terminological consistency, the key knowledge-centric concepts used in this article are summarized in Table~\ref{tab:kraken_terms}. These definitions provide a structured vocabulary for describing semantic efficiency, distortion, goal alignment, world modeling, intent representation, and distributed coordination. This shared terminology forms the foundation for the knowledge-centric framework developed in the remainder of the paper. By establishing common semantic abstractions for knowledge exchange and coordination, this vocabulary enables integrated communication, reasoning, and control across heterogeneous network entities.

\begin{table*}[h]
\centering
\caption{Key knowledge-centric networking concepts used in the Kraken architecture.}
\label{tab:kraken_terms}
\footnotesize
\renewcommand{\arraystretch}{1.3}
\setlength{\tabcolsep}{6pt}

\begin{tabular}{p{3.2cm} || p{14cm}}
\toprule
\textbf{Term} & \textbf{Definition} \\
\midrule

Semantic efficiency gain & Fraction of task-relevant information delivered (as estimated through a task-utility proxy) relative to total transmitted bits. \\

Semantic distortion & Loss incurred when raw observations are compressed into semantic representations. \\

Goal alignment error & Mismatch between local agent decisions and global service intents. \\

Goal-oriented metric & Performance indicator defined at the task or application level rather than conventional network-layer KPIs. \\

World model & Internal generative representation of network dynamics used for prediction and planning. \\

Generative Network Agent & Distributed network entity performing perception, memory, planning, and action via generative reasoning. \\

Knowledge object & Structured information unit enriched with context, confidence, and temporal validity. \\

Intent descriptor & Formal representation of service objectives and constraints guiding coordination. \\

Semantic negotiation & Iterative alignment of agent states and intents toward coordinated decisions. \\

Knowledge plane & Architectural layer maintaining shared knowledge, intents, and consistency across distributed agents. \\

\bottomrule
\end{tabular}
\end{table*}

Importantly, Kraken is presented as an architectural blueprint that synthesizes theoretical foundations with concrete engineering pathways toward deployable 6G intelligence. The authors are actively implementing the core components of this architecture including semantic-aware waveform processing, lightweight generative world models for edge devices, and knowledge graph synchronization mechanisms, with initial prototypes targeting Open RAN platforms and edge-cloud testbeds. Preliminary performance evaluations, reported throughout this paper, demonstrate feasibility across representative use cases: semantic compression ratios of 70-85\% for autonomous driving coordination, 10-20x bandwidth reduction for XR rendering, and up to 100:1 compression for industrial sensing while preserving task-relevant information. These early results, though obtained in controlled environments, validate the fundamental principles underlying Kraken and establish baselines for subsequent optimization. These prototype efforts illustrate how knowledge-centric networking principles can be translated into implementable system components within emerging 6G infrastructures.

The complete realization of Kraken as a unified network management system is necessarily a long-term endeavor, progressing through the phased deployment pathway outlined in Section~VIII. Each architectural component, the Infrastructure Plane's semantic-aware protocols, the Agent Plane's generative reasoning engines, and the Knowledge Plane's distributed graph substrate, presents distinct research challenges that will be addressed incrementally. Future publications will report technical contributions for each plane as implementations mature, including detailed performance characterization, standardization alignment, and large-scale validation. Kraken thus represents not a finished product but an evolving framework for collective intelligence, with this article establishing the foundational architecture against which subsequent experimental results will be measured.

The remainder of this article is organized as follows. Section II reviews related work and positions Kraken within prior research on semantic communication, AI-native networking, and goal-oriented control. Section III elaborates the conceptual shift from data-centric to knowledge-centric networking. Section IV presents the Kraken architecture in detail. Section V discusses enabling technologies for deployment. Section VI provides representative case studies. Section VII analyzes practicality and telco-grade requirements. Section VIII describes the evolutionary transition from 5G to Kraken-enabled 6G. Section IX outlines open challenges and future directions. Section X concludes the article.

\section{Related Work}

The Kraken architecture builds upon and synthesizes several research directions that have emerged over the past decade. In this section, we review the state of the art across six complementary domains, identify their limitations, and position Kraken relative to existing approaches. Table~I summarizes how Kraken addresses the key gaps observed in prior work.

\subsection{Semantic Communication: From Source Coding to Task-Aware Transmission}

The conceptual foundations of semantic communication can be traced to the Shannon--Weaver communication model \cite{shannon1998mathematical}, which distinguishes between technical (bit-level), semantic (meaning-level), and effectiveness (goal-level) layers of communication. Classical information theory has primarily addressed the technical layer, where rate–distortion theory establishes limits for signal reconstruction under fidelity constraints \cite{cover2006elements}. 

Recent research has begun to operationalize semantic communication by prioritizing task relevance over bit-level fidelity. The survey in \cite{9955525} formalized semantic and task-oriented communication as a framework in which transmitted representations are optimized for downstream inference or control objectives rather than exact signal recovery. Practical realizations have emerged through joint source–channel coding (JSCC) with deep neural networks. In particular, DeepJSCC \cite{8723589} demonstrated that learned end-to-end mappings can preserve perceptual and semantic content under bandwidth constraints. Subsequent extensions to video \cite{10.1007/s00530-025-02177-7} and speech \cite{9500590} confirmed similar advantages across modalities, while importance-aware transmission strategies further reduce communication overhead by allocating resources according to task contribution.

\noindent
\textbf{Limitations:}
Existing studies on semantic communication primarily focus on point-to-point links and single-task scenarios. Mechanisms for multi-agent semantic exchange, negotiation of conflicting interpretations, and alignment of distributed world models remain largely unexplored. In addition, most approaches assume static task objectives, whereas future 6G environments must adapt to evolving goals and contextual changes. Kraken addresses these limitations by embedding semantic communication within a distributed multi-agent architecture in which knowledge exchange is negotiated and explicitly aligned with system-level intents.

\subsection{AI/ML for Network Management: From Discriminative to Generative Intelligence}

Machine learning has been widely applied to network management tasks including resource allocation, scheduling, and handover optimization. Deep reinforcement learning (DRL) approaches have consistently outperformed rule-based heuristics in dynamic network environments \cite{8714026,8927868,Kumar2025}. Supervised learning and sequence models, such as recurrent neural networks and transformers, have also enabled accurate prediction of traffic dynamics and anomaly detection in network operations \cite{8581000,10867389}. Federated learning further enables distributed model training across network nodes while preserving data privacy \cite{https://doi.org/10.1002/ett.4458}.

Most of these approaches, however, remain fundamentally discriminative: models learn mappings from observations to predictions or decisions without explicitly modeling underlying system dynamics. More recently, generative models have been explored to capture distributions of network behavior. Generative adversarial networks have been used to synthesize realistic traffic traces for simulation and testing \cite{articlexx}, variational autoencoders learn compact latent representations for anomaly detection \cite{8802833}, and diffusion models generate rare or extreme traffic scenarios for protocol stress testing \cite{SIVAROOPAN2024110616}. Despite these advances, generative models are typically used offline for data generation or analysis rather than as online reasoning components embedded within network control loops.

\noindent
\textbf{Limitations:}
Current AI-driven network management primarily relies on predictive or reactive models that lack explicit representations of system dynamics and often degrade under distribution shifts. Generative world models, by contrast, capture underlying dynamics and uncertainty, enabling scenario reasoning and adaptive policy synthesis. Kraken introduces generative intelligence as an online capability through distributed Generative Network Agents that maintain and update world models within the network control loop, enabling adaptive decision-making rather than static learned mappings.
\begin{table*}[t]
\centering
\caption{Comparison of Kraken with representative research paradigms across six domains: 
semantic communication 
\cite{shannon1998mathematical,cover2006elements,8723589,9500590,9955525,10.1007/s00530-025-02177-7}, 
AI/ML-driven network management 
\cite{8714026,8927868,Kumar2025,8581000,10867389,https://doi.org/10.1002/ett.4458,articlexx,8802833,SIVAROOPAN2024110616}, 
goal- and task-oriented communication 
\cite{10110357,10644029,HexaX2023,9606667,10319671,10597087}, 
multi-agent networking and distributed control 
\cite{9580444,10089622,10848135,Oliehoek2016,3gpp23288,doi:https://doi.org/10.1002/9781119847083.ch4}, 
knowledge-graph-based networking 
\cite{rfc7285_full,10.1145/2656877.2656887,8368384,WANG2024124679,11143399,10.1145/1516533.1516536,10.1145/3502223.3502246}, 
and foundation-model-enabled telecommunications 
\cite{10466747,10588879,11162447,bilen2025knowledgedefinedtwinassistednetworkmanagement}.}
\label{tab:comparison}

\scriptsize
\renewcommand{\arraystretch}{1.2}

\begin{tabular}{p{1.5cm} p{2cm} p{2cm} p{2cm} p{2cm} p{2cm} p{2.9cm}}

\toprule

\textbf{Dimension} 
& \textbf{Semantic Comm.} 
& \textbf{AI/ML Management} 
& \textbf{Goal-Oriented} 
& \textbf{Multi-Agent} 
& \textbf{Knowledge Graphs} 
& \textbf{Kraken (Proposed Multi-Plane Architecture)} \\
\midrule
\midrule

\textbf{Intelligence Type} 
& Task-aware encoding 
& Discriminative learning 
& Objective-driven optimization 
& Reinforcement learning 
& Graph-based reasoning 
& Distributed generative reasoning \\

\hline

\textbf{Representation} 
& Compressed features 
& Raw metrics / KPIs 
& Task-level metrics 
& State vectors 
& Fixed ontologies 
& Knowledge objects (facts, intents, models, uncertainty, temporal validity) \\

\hline

\textbf{Coordination} 
& Point-to-point 
& Centralized analytics 
& Implicit intent translation 
& Emergent coordination via reward signals 
& Query-based access 
& Multi-agent negotiation with world-model alignment \\

\hline

\textbf{Objective} 
& Task-relevant information delivery 
& QoS/KPI optimization 
& Task success probability 
& Local reward maximization 
& Relation discovery 
& Goal/task-level outcomes with formal alignment \\

\hline

\textbf{Adaptation} 
& Fixed encoding 
& Model retraining 
& Policy refinement 
& Online learning 
& Periodic graph update 
& Continuous world-model update under uncertainty \\

\hline

\textbf{Deployment} 
& Edge links 
& Core/RAN modules 
& Service-layer optimization 
& Simulation/testbeds 
& Management plane 
& Full-stack integration with phased migration \\

\hline

\textbf{Key Limitation} 
& Single-task focus, no negotiation 
& Reactive, no world model 
& Architectural underspecification 
& No structured knowledge exchange 
& Static, non-negotiated knowledge 
& --- \\

\bottomrule

\end{tabular}

\end{table*}
\subsection{Goal-Oriented and Task-Oriented Communication}

The recognition that communication objectives extend beyond reliable bit delivery has motivated the emerging paradigm of task- and goal-oriented communication. In task-oriented communication, transmitted information is formulated as the minimal sufficient statistic required to achieve a specific inference or control objective \cite{10110357}. More broadly, goal-oriented communication extends this principle to system-level objectives, where communication and control are jointly optimized to maximize task effectiveness rather than intermediate QoS metrics. The 6G vision has increasingly embraced this shift. Zhou \textit{et al.} \cite{10644029} identify semantic and goal-oriented communication as key enablers of intelligent wireless systems, while the Hexa-X initiative highlights task effectiveness as an important performance indicator beyond conventional throughput–latency metrics \cite{HexaX2023}.

Several practical realizations have begun to emerge. Task-oriented source coding for edge inference optimizes encoded representations for downstream decision accuracy rather than reconstruction fidelity \cite{9606667}. Semantic-aware resource allocation incorporates the relevance of each data flow to application objectives into scheduling decisions \cite{10319671}. More recently, goal-oriented optimization frameworks formulate network control as maximizing task success probability under resource constraints \cite{10597087}. Collectively, these works demonstrate the feasibility and potential benefits of communication systems designed around task relevance and the satisfaction of intent.

\noindent
\textbf{Limitations:}
Existing goal- and task-oriented communication research primarily defines the objective paradigm but provides limited architectural guidance for distributed network systems. The mechanisms through which semantic representations, reasoning processes, and goal alignment interact across heterogeneous network entities remain largely unspecified. Kraken addresses this limitation by embedding goal-oriented objectives within the Knowledge Plane, translating them into coordination constraints and operationalizing them through generative planning in distributed agents.

\subsection{Multi-Agent Systems for Telecommunications}

Distributed decision-making in telecommunications has been widely studied through multi-agent reinforcement learning (MARL) and decentralized control frameworks. Applications include distributed spectrum sharing \cite{9580444}, multi-cell interference coordination \cite{10089622}, and edge computing orchestration \cite{10848135}, where multiple agents learn policies that balance local performance with network-wide efficiency. Centralized training with decentralized execution (CTDE) paradigms \cite{Oliehoek2016} further enable agents to learn coordinated behaviors through shared training while preserving distributed autonomy during deployment.

Telecommunications architectures have also begun incorporating elements of distributed intelligence. The 3GPP Network Data Analytics Function (NWDAF) \cite{3gpp23288} provides analytics services to network functions, acting as a logically centralized data-driven intelligence component. Similarly, the O-RAN architecture introduces the near-real-time RAN Intelligent Controller (RIC), which hosts xApps for localized control, and the non-real-time RIC hosting rApps for longer-timescale optimization \cite{doi:https://doi.org/10.1002/9781119847083.ch4}. These developments represent important steps toward intelligent and distributed network control. However, they remain largely application-specific extensions rather than a unified cognitive framework spanning network layers and operational domains.

\noindent
\textbf{Limitations:}
Existing multi-agent networking approaches typically treat agents as independent learners optimizing local reward functions, with coordination emerging through shared observations or predefined signaling interfaces. As a result, exchanged information is usually limited to measurements or engineered features rather than structured semantic abstractions enriched with confidence, temporal validity, and intent context. Kraken advances beyond this paradigm by introducing a distributed Knowledge Plane in which agents exchange and align world models rather than raw observations, thereby enabling a negotiated, shared understanding of network state and coordinated goal-oriented decision-making across heterogeneous domains.

\subsection{Knowledge Graphs and Semantics in Networking}

Semantic abstractions and structured knowledge representations have been explored in networking through several architectural and management paradigms. The IETF Application-Layer Traffic Optimization (ALTO) protocol \cite{rfc7285_full} exposes network topology and cost information to applications, enabling informed path selection. However, ALTO relies on relatively static, coarse-grained abstractions rather than on context-aware semantic reasoning. Named Data Networking (NDN) \cite{10.1145/2656877.2656887} shifts addressing from host locations to content names, introducing an information-centric communication model, yet it does not incorporate structured reasoning about content relevance, task contribution, or intent alignment.

Semantic routing proposals \cite{8368384} enrich packets with ontology-based metadata to guide forwarding decisions, particularly in IoT service discovery scenarios. While these approaches attach semantic labels to flows, such metadata are typically defined during session establishment and remain fixed, without mechanisms for dynamic negotiation or continuous contextual adaptation.

Knowledge graphs have also been adopted in network management systems. They have been used for root cause analysis \cite{WANG2024124679}, automated configuration validation \cite{11143399}, and semantics-based service composition \cite{10.1145/1516533.1516536}. More recent work integrates knowledge graphs with machine learning to support explainable anomaly detection and structural reasoning over network events \cite{10.1145/3502223.3502246}. These studies demonstrate the value of structured relational representations for capturing dependencies among network entities, services, and performance indicators.

\noindent
\textbf{Limitations:}
Despite their expressive power, prior approaches treat knowledge graphs primarily as static or periodically updated repositories queried by centralized management systems. Knowledge elements are typically represented as deterministic facts, lacking explicit temporal validity, confidence estimation, provenance tracking, or negotiation mechanisms. Consequently, they support querying and inference but not dynamic alignment across distributed agents with potentially conflicting observations. Kraken advances beyond this paradigm by defining a distributed Knowledge Plane in which knowledge is represented as enriched semantic objects that incorporate contextual state, uncertainty, and intent attributes. Rather than serving as a passive database, the Knowledge Plane functions as a dynamic coordination substrate through which agents align world models and negotiate shared understanding of network state under uncertainty.

\subsection{Foundation Models for Telecommunications}

The rapid advancement of large-scale foundation models, including large language models (LLMs) and multimodal architectures pretrained on massive datasets, has stimulated increasing interest in their application to telecommunications systems. These models demonstrate strong generalization, enabling adaptation to diverse downstream tasks with minimal fine-tuning. Recent studies have explored their role in network operations. NetGPT \cite{10466747} proposes an AI-native architecture that leverages LLMs to assist network provisioning and diagnostics through natural-language interaction. Similarly, LLM-based intent translation frameworks \cite{10588879} map natural-language operator goals into structured network configurations, supporting intent-driven network management paradigms.

Foundation models have also been explored in conjunction with network digital twins. Predictive twin frameworks \cite{11162447} and knowledge-defined twin-assisted architectures \cite{bilen2025knowledgedefinedtwinassistednetworkmanagement} employ large-scale generative models to synthesize failure scenarios, anticipate hardware degradation, and generate synthetic traffic patterns for proactive network management. These approaches highlight the potential of pretrained generative models to enhance simulation, diagnostics, and planning in complex 6G environments.

\noindent
\textbf{Limitations:}
Despite these promising developments, most current applications treat foundation models primarily as external tools for human interaction, intent parsing, or offline scenario generation rather than as integrated components of real-time network control. As a result, they do not fully address the stringent requirements of telecommunications systems, including predictable inference latency, bounded hallucination risk, verifiable reasoning paths, and adversarial robustness. Kraken therefore does not assume that contemporary foundation models can be directly deployed within critical control loops. Instead, it defines an engineering trajectory toward Telco-Grade Foundation Models, positioning them as epistemic priors within the Knowledge Plane and embedding them within a constrained, hierarchical, and verifiable control framework rather than granting them direct autonomous authority.

\subsection{Comparison with 6G Architectural Proposals}

Several comprehensive architectural visions for 6G have emerged in recent years, outlining performance targets, enabling technologies, and system-level aspirations. The Hexa-X initiative articulates a European vision for 6G centered on sustainability, trustworthiness, and extreme performance while defining key use cases and system requirements. However, it primarily specifies capability goals and architectural principles rather than a concrete, knowledge-centric control framework that integrates semantic communication, distributed reasoning, and goal alignment within a unified operational structure. Similarly, the ITU-R IMT-2030 framework \cite{ITUR2024_IMT2030} establishes usage scenarios, capability requirements, and high-level functional trends for 6G systems. While it acknowledges the importance of AI and intelligent automation, it does not detail how artificial intelligence components are structurally embedded into the control architecture, nor how distributed reasoning and coordination mechanisms can be operationalized across heterogeneous network domains.

Closer to our technical direction, Hoydis \textit{et al.} \cite{Hoydis2020TowardA6} advocate an AI-native air interface that integrates deep learning into physical-layer processing and waveform design. This line of research represents an important step toward AI-driven physical-layer optimization. However, its scope remains focused on link-level adaptation and does not address multi-agent coordination, semantic negotiation, or cross-layer knowledge alignment at the network scale. Talwar \textit{et al.} \cite{9665431} envision 6G as enabling a form of distributed intelligence across devices and infrastructure, conceptualizing the network as a distributed brain. While this perspective aligns philosophically with collective intelligence, it remains primarily visionary and does not specify the architectural mechanisms required for structured knowledge exchange, generative world modeling, or goal-oriented coordination.

\noindent
\textbf{Limitations:}
Existing 6G architectural proposals articulate performance ambitions and intelligent capabilities but generally stop short of defining a concrete knowledge-centric control architecture. They either emphasize requirements and high-level principles, focus on specific layers such as the physical interface, or describe distributed intelligence conceptually without defining operational mechanisms for semantic representation, generative reasoning, and negotiated alignment across agents. Kraken addresses this gap by proposing an explicit multi-plane architecture in which a distributed Knowledge Plane, Generative Network Agents, and a semantic-aware Infrastructure Plane are integrated in a structured manner. Rather than treating AI as an auxiliary enhancement or a layer-specific optimization tool, Kraken embeds intelligence as a first-class architectural construct designed for coordinated, goal-aligned, and evolutionarily deployable 6G systems.
\subsection{Summary and Positioning}

The preceding subsections show that substantial advances have been achieved across semantic communication, AI-driven network management, goal-oriented optimization, multi-agent coordination, knowledge graphs, foundation models, and emerging 6G architectural visions. However, these directions have largely progressed independently without converging into a unified architectural control framework. Table I synthesizes this comparison and illustrates how Kraken consolidates these strands within a coherent and standards-aligned knowledge-centric architecture. The distinguishing characteristics of Kraken are summarized as follows:

\begin{itemize}

\item \textit{Cross-layer semantic integration:}  
Existing 6G proposals often introduce intelligence at specific layers, such as physical-layer learning or intent-based management. Kraken, by contrast, embeds semantic awareness vertically across the protocol stack. Semantic representations propagate from physical-layer processing and scheduling to application-level intent formulation, enabling communication, inference, and coordination within a unified multi-plane architecture.

\item \textit{Distributed generative agents with explicit world models:}  
Most AI-enabled networking frameworks rely on predictive or discriminative models that map observations to actions. Kraken introduces Generative Network Agents that maintain explicit world models of network dynamics, enabling structured reasoning, scenario synthesis, and adaptive planning under uncertainty beyond reactive optimization.

\item \textit{Formal goal-oriented alignment mechanisms:}  
Although intent-driven networking is widely discussed in 6G visions, goal alignment is often expressed qualitatively. Kraken operationalizes goal-oriented optimization through explicit objective representations in the Knowledge Plane, translating service intents into coordination constraints across distributed agents.

\item \textit{Knowledge-centric abstraction for coordination:}  
Prior systems primarily exchange measurements, engineered features, or predefined metadata. Kraken elevates knowledge objects enriched with contextual information, confidence estimates, temporal validity, and provenance as the primary abstraction for distributed coordination, enabling reasoning under uncertainty and reconciliation of divergent observations across heterogeneous entities.

\item \textit{Engineering-grounded and evolutionarily deployable architecture:}  
Many 6G proposals assume disruptive redesigns or treat artificial intelligence as an auxiliary capability. Kraken instead incorporates complexity–practicality trade-offs, defines requirements for Telco-Grade Foundation Models, and embeds generative reasoning within a constrained hierarchy. It further defines a phased transition from existing 5G systems toward knowledge-centric 6G deployment, ensuring backward compatibility and operational continuity.

\end{itemize}

Collectively, Kraken does not merely extend existing AI-native or semantic-enhanced networking architectures. It introduces a multi-plane control paradigm in which semantic communication, generative reasoning, and goal-oriented optimization are jointly embedded within a distributed knowledge-centric framework. By treating knowledge as a first-class abstraction and enabling world-model alignment across heterogeneous entities, Kraken reframes network intelligence as an engineered and deployable foundation for realistic 6G systems.


\begin{figure*}[h]
	\centering
	\resizebox{\textwidth}{!}{%
		\begin{tikzpicture}[
			node distance=3.2cm,
			start chain=going right,
			block/.style={
				on chain, rectangle, 
				draw=blue!50!black!30, 
				top color=white, bottom color=blue!10,
				text width=3.5cm,
				minimum height=2.2cm, 
				align=center,
				rounded corners=10pt, 
				thick,
				drop shadow={opacity=0.15, shadow xshift=2pt, shadow yshift=-2pt},
				font=\sffamily\normalsize,
				inner sep=6pt
			},
			transition/.style={
				-{Stealth[scale=1.5]}, 
				line width=1.5pt, 
				color=gray!60,
				shorten >=6pt, shorten <=6pt
			},
			shift_label/.style={
				font=\sffamily\small\bfseries,
				align=center,
				yshift=12pt,
				text=blue!40!black
			},
			krakenbox/.style={
				rectangle,
				draw=blue!80!black!40,
				fill=blue!5,
				fill opacity=0.2,
				rounded corners=20pt,
				line width=2pt,
				inner sep=25pt,
				dash pattern=on 10pt off 3pt
			}
			]
			
			\node[block, bottom color=gray!15, draw=gray!40] (data) {
				\textbf{\large Data-Centric Networking}\\ \vspace{2pt}
				\small Raw Measurements, Bits \\ \& Protocol Headers
			};
			
			\node[block] (semantic) {
				\textbf{\large Semantic Communication}\\ \vspace{2pt}
				\small Knowledge Objects \\ Structured Abstractions
			};
			
			\node[block] (gen) {
				\textbf{\large Generative Reasoning}\\ \vspace{2pt}
				\small World Models \\ Scenario Synthesis
			};
			
			\node[block] (goal) {
				\textbf{\large Goal-Oriented Optimization}\\ \vspace{2pt}
				\small Task Success \\ \& Intent Satisfaction
			};
			
			\node[block, bottom color=orange!20, draw=orange!50, text=orange!50!black] (coll) {
				\textbf{\large Collective Intelligence}\\ \vspace{2pt}
				\small Distributed Multi-Agent \\ Emergent Intelligence
			};
			
			\draw[transition] (data) -- (semantic) 
			node[midway, above, shift_label] {Representation\\Transformation};
			
			\draw[transition] (semantic) -- (gen) 
			node[midway, above, shift_label] {Reasoning\\Integration};
			
			\draw[transition] (gen) -- (goal) 
			node[midway, above, shift_label] {Objective\\Alignment};
			
			\draw[transition] (goal) -- (coll) 
			node[midway, above, shift_label] {Distributed\\Scaling};
			
			\begin{scope}[on background layer]
				\node[krakenbox, fit=(data)(semantic)(gen)(goal)(coll)] (kraken) {};
			\end{scope}
			
			\node[
			fill=white, 
			draw=blue!80!black, 
			thick, 
			rounded corners=5pt,
			font=\sffamily\Large\bfseries, 
			text=blue!80!black,
			inner sep=11pt,
			anchor=center
			] at (kraken.north)
			{Evolutionary Roadmap to Collective Intelligence};
			
		\end{tikzpicture}%
	}
	\caption{Evolution of networking from a data-centric approach toward distributed collective intelligence in 6G systems through semantic abstraction, generative reasoning, and goal-oriented coordination.}
	\label{fig:evolution_roadmap}
\end{figure*}
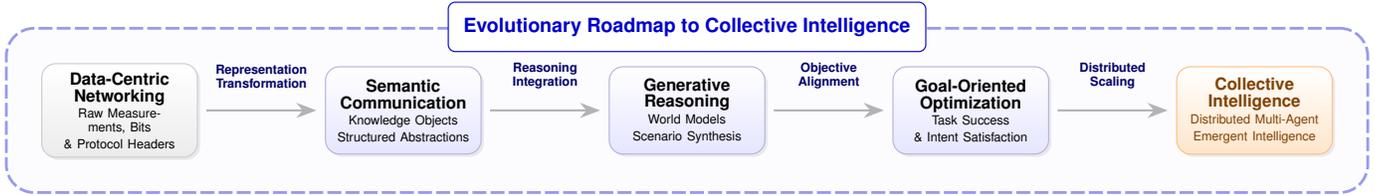

\begin{table*}[h]
	\centering
	\caption{Operational requirements and evaluation dimensions associated with the transition from data-centric networking to collective intelligence in 6G systems.}
	\label{tab:kraken_requirements}
	\scriptsize
	\renewcommand{\arraystretch}{1.3}
	
	\begin{tabular}{p{2.8cm} p{7.3cm} p{6.7cm}}
		
		\toprule
		
		\textbf{Architecture stage} 
		& \textbf{Representation and capability shift} 
		& \textbf{Network implication} \\
		
		\midrule
		
		Data-Centric Networking 
		& Raw measurements, protocol headers, and bit-level transport 
		& Communication optimized for throughput and reliability without task awareness \\
		
		Semantic communication 
		& Knowledge-centric structured abstractions and semantic representations 
		& Relevance-aware transmission and reduced communication entropy \\
		
		Generative reasoning 
		& Internal world models and predictive scenario synthesis 
		& Proactive network management and anticipatory adaptation \\
		
		Goal-oriented optimization 
		& Task success and intent satisfaction integrated into control objectives 
		& Resource allocation aligned with application-level outcomes \\
		
		Collective intelligence 
		& Distributed multi-agent coordination over a shared knowledge plane 
		& Emergent global behavior and scalable system-level intelligence \\
		
		\bottomrule
		
	\end{tabular}
\end{table*}

\section{From Data-Centric Networking to Collective Intelligence}

The transition from conventional data-centric networking toward distributed collective intelligence is not a single-step evolution but a multi-dimensional transformation, as illustrated in Fig.~\ref{fig:evolution_roadmap}. It requires rethinking how information is represented, how intelligence is embedded within the network, and how system-level objectives are defined and coordinated. In the following, we articulate the key dimensions of this transition, \textit{semantic communication}, \textit{generative reasoning}, and \textit{goal-oriented coordination}, and show how their systematic integration enables collective intelligence in 6G systems.

\subsection{From Data-Centric to Knowledge-Centric Semantic Networking}

Conventional network architectures are inherently data-centric. Network entities exchange raw measurements, protocol headers, and payload bits, while interpretation of meaning is deferred to upper-layer applications or external services. Resource allocation mechanisms operate on traffic volume and channel conditions without awareness of how transmitted information contributes to system-level objectives. In emerging 6G scenarios, this separation becomes increasingly inefficient. Many applications are not sensitive to bit-level fidelity but to the correctness of inferred states, successful task execution, or timely decision support. The utility of communication therefore depends less on accurate symbol reconstruction and more on the relevance of exchanged information.

This observation motivates a transition toward knowledge-centric networking, in which nodes exchange structured knowledge objects such as inferred states, predicted events, intent representations, causal relations, and associated confidence measures. Data corresponds to raw signal observations, whereas knowledge represents compressed, structured, and task-relevant abstractions derived from those observations \cite{11357882}. By transmitting knowledge rather than raw data, the network reduces communication entropy while increasing decision relevance.

Recent advances in semantic communication support this shift by reformulating the design objective from bit-level reliability to task-level effectiveness. Importance-aware encoding mechanisms demonstrate that relevance-driven transmission can significantly reduce bandwidth consumption while improving robustness under constrained channel conditions. Moreover, modern foundation models enable multimodal semantic fusion and joint reasoning over heterogeneous inputs, including channel state information, traffic statistics, geospatial context, event logs, and application descriptors. This unified representation allows the network to capture latent cross-modal correlations that remain invisible to unimodal processing pipelines.

By embedding semantic awareness and multimodal reasoning into the communication process, the network evolves from a passive transport substrate into an active participant in system-level decision-making. However, semantic exchange alone is insufficient. Once information is represented at higher abstraction levels, the network must be capable of reasoning over that knowledge, generalizing to unseen conditions, and synthesizing adaptive operational strategies. This requirement naturally motivates integrating generative reasoning.

\subsection{Generative Reasoning for Semantic Network Management}

Most existing AI-enabled network management solutions rely on discriminative models trained on historical datasets. While effective under relatively stationary conditions, such models do not construct internal generative representations of system dynamics and therefore exhibit limited adaptability in novel scenarios.

Generative reasoning introduces a fundamentally different capability. By modeling data distributions and environmental structure, generative systems construct internal world models that approximate network dynamics. These models enable agents to synthesize unseen configurations, simulate alternative system trajectories, and anticipate future states. Within network management, this capability supports policy generation for unfamiliar traffic patterns, proactive optimization through scenario exploration, and structured translation between high-level service intents and low-level resource configurations.

Beyond distribution modeling, large-scale foundation models demonstrate emergent reasoning properties that are particularly relevant for 6G environments. In-context adaptation enables rapid generalization to new interference patterns or service profiles without explicit retraining. Instruction-level reasoning supports intent-to-policy translation, while stepwise reasoning mechanisms improve interpretability by exposing the rationale behind generated decisions. Through continuous interaction and feedback, generative agents refine their internal world models, enabling adaptation under non-stationarity and cross-domain coupling. In this sense, generative reasoning transforms the network from a reactive optimizer into a predictive and anticipatory system.

Yet reasoning capability alone does not determine what should be optimized. Without principled objective alignment, generated policies may optimize surrogate metrics disconnected from actual application outcomes. Intelligence must therefore be grounded in a goal-oriented optimization framework.

\subsection{From Semantics to Goal-Oriented Networks}

Traditional wireless optimization is formulated around quality-of-service metrics such as throughput, latency, packet loss, and spectral efficiency. While these metrics remain essential performance indicators, they often serve as intermediate variables rather than ultimate objectives in emerging 6G applications. In cooperative autonomous driving, the primary objective is safe maneuver coordination. In immersive XR, perceptual consistency and synchronization dominate raw bitrate requirements. In industrial automation, deterministic actuation reliability supersedes average delay minimization. These examples illustrate that application-level outcomes must directly shape communication and resource allocation decisions.

Goal-oriented networking formalizes this perspective by incorporating task success probability, semantic distortion, intent satisfaction, and outcome reliability into the control loop. Communication, computation, and coordination layers become tightly coupled through continuous semantic feedback. Generative agents operating within this framework no longer optimize abstract QoS metrics alone, but synthesize actions that maximize task-level performance under uncertainty.

However, as networks scale in density, heterogeneity, and domain diversity, centralized goal optimization becomes computationally impractical and structurally fragile. Ensuring alignment between local decisions and global objectives therefore requires a distributed realization of goal-oriented intelligence.

\subsection{From Goal Alignment to Distributed Collective Intelligence}

The scale and heterogeneity envisioned for 6G necessitate inherently distributed intelligence. Centralized control architectures face scalability constraints, signaling overhead, and vulnerability to single points of failure. Collective intelligence provides an alternative paradigm in which system-level capabilities emerge from interactions among autonomous agents.

We consider a multi-agent architecture in which edge devices, access points, and core network entities operate as collaborative generative agents. Each agent maintains a local world model representing a compressed view of the network state relevant to its operational scope. Rather than exchanging raw measurements, agents communicate through a logical knowledge plane that carries structured semantic representations, including predictions, intentions, confidence scores, and coordination requests. Communication thus becomes a process of negotiation and model alignment rather than a simple reporting process.

To preserve scalability while enabling global optimization, agents can be organized hierarchically. Edge agents embedded in user equipment or sensors perform low-latency local reasoning. Infrastructure agents located at base stations or edge servers maintain broader contextual awareness and coordinate neighboring nodes. Domain-level agents residing in cloud or regional controllers synthesize cross-domain policies and long-term strategies. This hierarchical yet autonomous structure ensures that latency-critical decisions are resolved locally, while context-intensive coordination is handled at higher abstraction levels.

Within such a distributed knowledge plane, intelligence emerges through continuous interaction among semantic representations, generative world models, and goal-oriented coordination mechanisms across heterogeneous network entities. Rather than exchanging raw measurements, agents share compact knowledge objects that encode inferred states, predicted dynamics, and intent descriptors, along with associated confidence and temporal validity. This structured exchange enables local agents to adapt to their environments while remaining aligned with system-wide objectives through iterative knowledge refinement and intent negotiation.

\begin{figure*}[h]
\centering
\includegraphics[width=0.74\linewidth]{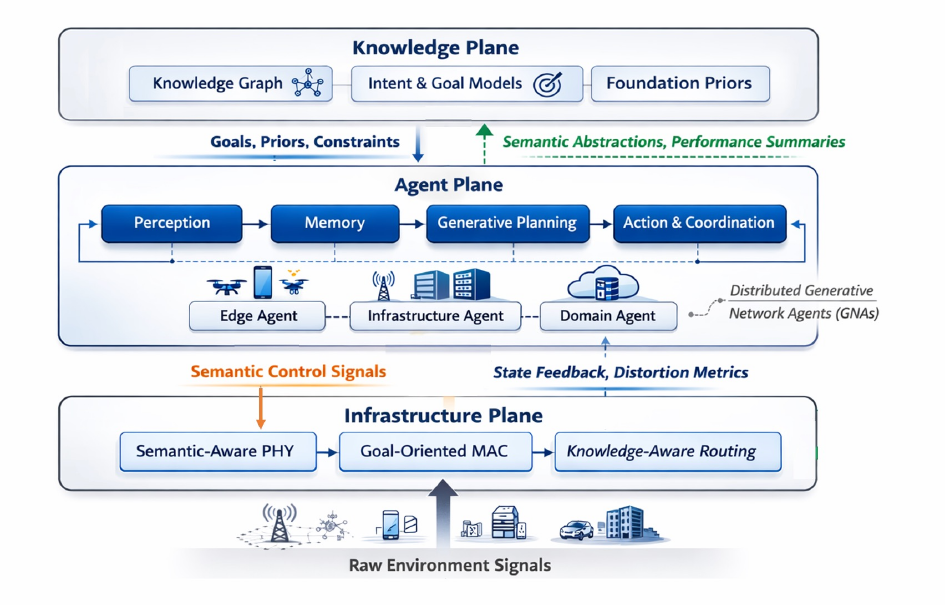}
\caption{The KRAKEN three-plane architecture showing the Infrastructure Plane (PHY/MAC/NET layers with semantic awareness), Agent Plane (distributed Generative Network Agents with perception-memory-planning-action cycles), and Knowledge Plane (semantic substrate, foundation model priors, and goal governance). Arrows indicate bidirectional information flow and closed-loop control.}
\label{arc}
\end{figure*}

As a result, functions traditionally treated as isolated control problems, such as routing, scheduling, resource allocation, and service orchestration, become coupled through semantic context and predictive reasoning. Semantic routing, multimodal fusion, anticipatory optimization, and cross-layer orchestration thus arise as emergent properties of coordinated multi-agent behavior rather than centrally imposed policies. The network evolves from a reactive transport substrate into a distributed cognitive system capable of anticipation, alignment, and collective adaptation under uncertainty.

This transition from data-centric communication toward collective intelligence is summarized conceptually in Fig.~\ref{fig:evolution_roadmap}, while its operational requirements and evaluation dimensions are synthesized in Table~\ref{tab:kraken_requirements}. Together, they illustrate how semantic abstraction, generative reasoning, and goal alignment reshape both the functional architecture and measurable performance criteria of 6G networks, linking conceptual intelligence layers with deployable mechanisms and verifiable metrics. The architectural realization of this integration is presented in the next section through the Kraken framework.

\section{Kraken: A Knowledge-Centric Architecture for 6G Collective Intelligence}
Kraken is designed as a hierarchical multi-plane architecture that transforms conventional 6G infrastructures into a distributed collective intelligence system. Rather than introducing an external intelligence overlay, Kraken restructures the network architecture itself by embedding semantic awareness, generative reasoning, and goal-oriented coordination across tightly coupled functional planes, as illustrated in Fig.~\ref{arc}. The architecture is organized into three vertically integrated planes: 
(i) the Infrastructure Plane, 
(ii) the Agent Plane, and 
(iii) the Knowledge Plane. 
Each plane extends classical 6G mechanisms while preserving backward compatibility and reinterpreting their functions within a knowledge-centric closed control loop. Intelligence therefore does not reside in a single centralized entity but emerges from structured interaction across planes, forming a distributed collective intelligence layer.

\begin{figure*}[h]
\centering
\includegraphics[width=\linewidth]{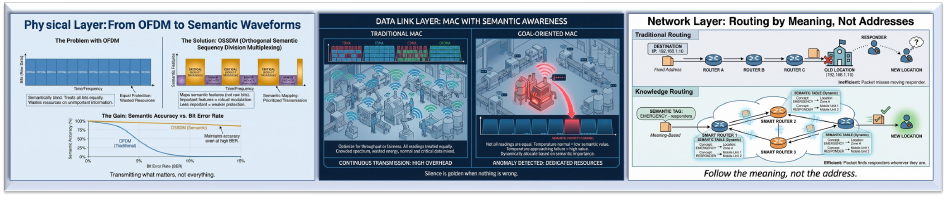}
\caption{Infrastructure Plane of the Kraken architecture with semantic-aware Physical, Data Link, and Network layers.}
\label{f2}
\end{figure*}

\vspace{-0.1in}
\subsection{Kraken Infrastructure Plane}
At the foundation of Kraken lies the Infrastructure Plane, which incorporates advanced 6G radio technologies while embedding semantic awareness into waveform design, medium access control, and routing decisions. This plane maintains compatibility with OFDM-based multicarrier transmission, massive MIMO, adaptive modulation and coding, and cross-layer scheduling mechanisms envisioned for 6G deployments. Unlike conventional systems that optimize exclusively for bit-level reliability and spectral efficiency, Kraken introduces semantic importance as an additional control dimension. Under this paradigm, the classical layered communication stack evolves into a semantically augmented infrastructure, as detailed below.

\subsubsection{Physical Layer -- Semantic-Aware Waveforms}
In classical OFDM-based systems, subcarriers are allocated according to channel state information and traffic demand. In Kraken, subcarrier allocation, symbol energy distribution, and coding redundancy are additionally weighted by semantic contribution scores provided by the Agent and Knowledge planes. Task-critical semantic tokens are assigned higher transmission reliability through mechanisms such as increased power allocation, stronger coding schemes, or prioritized subcarrier mapping, whereas low-impact information may be transmitted with lighter protection. This importance-aware transmission strategy generalizes semantic subcarrier prioritization and transforms the physical layer from a purely channel-centric module into a task-aware communication substrate. Importantly, the physical layer does not interpret semantic meaning; it executes priority signals computed by higher planes. It therefore remains fully compatible with existing multicarrier implementations while acting as an executor of higher-level coordination policies, as summarized in Fig.~\ref{f2}.

\subsubsection{Data Link Layer -- Goal-Oriented Multiple Access}
At the Medium Access Control (MAC) layer, scheduling is extended to operate on semantically annotated traffic flows rather than homogeneous packet queues. In conventional MAC design, resource blocks are allocated based on buffer occupancy, channel quality indicators, and fairness constraints.  In Kraken, scheduling decisions additionally incorporate semantic priority levels, task deadlines, and estimated goal-contribution metrics computed by upper planes. Each traffic flow is associated with a dynamic semantic weight reflecting its impact on task-level performance. The MAC scheduler therefore performs intent-aware prioritization, adjusting time–frequency resource allocation to reduce semantic distortion rather than merely maximizing throughput. Collision resolution, retransmission control, and hybrid ARQ mechanisms are similarly adapted. Retransmission policies depend not only on packet error rates but also on the semantic relevance of lost information. Semantic weights are continuously updated through feedback from the Agent and Knowledge planes, ensuring that link-layer actions remain aligned with evolving system-level objectives.

\subsubsection{Network Layer -- Knowledge-Aware Routing}
At the network layer, routing decisions incorporate semantic descriptors attached to packets, as illustrated in Fig.~\ref{f2}. Instead of selecting paths solely based on link quality indicators or hop count, forwarding mechanisms consider task identifiers, latency sensitivity, reliability requirements, and goal-contribution scores. Path selection thus becomes context-sensitive: routes that better preserve semantic fidelity or reduce task-level distortion are preferred over purely throughput-oriented alternatives. In multi-domain environments, routing decisions may additionally leverage predicted congestion states provided by generative agents, enabling anticipatory path selection. 

Despite these enhancements, the Infrastructure Plane does not perform semantic reasoning or interpret application-level objectives. It executes reliability and priority signals generated by higher planes. Semantic abstraction, multimodal fusion, and structured knowledge representation are therefore defined above this layer. This separation of concerns motivates the introduction of the Agent Plane, where heterogeneous raw data streams are transformed into unified knowledge objects that can be processed by generative reasoning mechanisms.

\subsection{Kraken Agent Plane}

The operational intelligence of Kraken is realized within the Agent Plane, where distributed Generative Network Agents (GNAs) embed semantic reasoning, predictive modeling, and goal-oriented coordination directly into network nodes. Unlike conventional control entities that execute predefined optimization routines, GNAs follow a structured cognitive architecture inspired by autonomous decision systems. Rather than separating sensing, optimization, and control into isolated modules, each agent integrates these functions into a unified reasoning loop that continuously interacts with the underlying infrastructure.

While all GNAs share a common cognitive architecture, they are deployed hierarchically to ensure scalability and functional differentiation. Edge-level agents reside within user equipment or sensing devices and perform low-latency local reasoning under strict resource constraints. Infrastructure-level agents operate at base stations and edge servers, maintaining broader contextual awareness across multiple edge nodes. Domain-level agents, typically hosted in regional or cloud environments, maintain long-term global world models and synthesize cross-domain coordination strategies. This hierarchical organization preserves local autonomy while enabling global objective alignment without requiring centralized optimization.

The internal cognitive structure of each agent is organized around four tightly coupled components: \emph{perception}, \emph{memory}, \emph{planning}, and \emph{action}. Together, these components form a closed reasoning cycle that transforms raw infrastructure signals into coordinated, goal-aligned control decisions.

\begin{itemize}

\item \textbf{Perception:}
The perception component transforms multi-layer infrastructure signals into structured semantic representations. Physical-layer indicators such as SINR, modulation and coding schemes, and channel quality reports; MAC-layer scheduling states; routing metadata; geospatial context; and application-level descriptors are jointly processed through multimodal fusion mechanisms. Rather than forwarding raw measurements to static optimization routines, the agent constructs structured semantic objects encapsulating inferred system state, contextual attributes, intent relevance, confidence levels, and temporal validity.
Semantic distortion is quantified as the divergence between the abstracted semantic representation and the underlying physical conditions. Perception therefore becomes an active semantic construction process rather than a passive sensing mechanism.
\item \textbf{Memory:}
Perceived semantic objects are integrated into the agent’s memory structure, implemented as a distributed knowledge graph combined with internal latent state embeddings. Through entity alignment, relation instantiation, and temporal linking, locally observed states become embedded within a relational semantic context that can be shared across agents through the Knowledge Plane. This memory layer captures historical evolution, propagates confidence estimates, models semantic affinity relationships, and enforces cross-agent consistency constraints. Semantic affinity weights quantify how strongly local decisions contribute to system-level objectives, directly influencing scheduling priorities, routing preferences, and coordination intensity.
\item \textbf{Planning:}
Planning is driven by the generative world model residing at the core of the agent. Rather than learning static input–output mappings, the generative core models the joint dynamics of network states, control actions, and task outcomes through a parametric high-dimensional world model, typically implemented as a latent-state transition architecture. 
This model learns the conditional relationship among the system's latent semantic state, the agent’s actions, and the resulting task-relevant reward or distortion signal. Architecturally, it can be realized using Transformer-based or MLP-based recurrent dynamics predictors trained on historical telemetry and operational feedback. Using this world representation, the agent synthesizes hypothetical scenarios, anticipates semantic distortion evolution, evaluates coordination strategies, and predicts distribution shifts in traffic demand or user mobility. The model enables the agent to simulate future trajectories (“dreaming”), evaluate counterfactual actions, and reason about uncertainty by sampling from its latent distribution. In this way, raw perception is transformed into structured foresight for adaptive decision-making. Unlike conventional model-predictive control, which relies on a known or manually specified system model, or standard multi-agent reinforcement learning approaches that learn reactive policies, the GNA world model is a learned latent representation of the environment's dynamics. This enables reasoning over partially observed states and supports generalization to previously unseen network conditions, which is essential for non-stationary 6G environments. To operationalize goal-oriented coordination under uncertainty, each agent formulates its planning problem as a constrained optimization task. The objective is to maximize local task success probability while respecting global intent constraints defined by the Knowledge Plane, such as maximum semantic distortion or resource consumption limits. This problem is solved using a Lagrangian coordination framework, in which the local objective is augmented with penalty terms for constraint violations, weighted by learnable dual variables (Lagrange multipliers). These multipliers are periodically updated by higher-level agents based on global system feedback, allowing the network to reach a dynamic equilibrium between local autonomy and global alignment. This formulation decomposes a complex multi-agent optimization problem into a set of decentralized yet loosely coupled subproblems with formal convergence guarantees.

\item \textbf{Action and Coordination:}
The outcome of planning is translated into coordinated control actions at the PHY, MAC, and routing layers. Actions reflect both local predictions and alignment with global intent constraints. Rather than acting independently, agents exchange compressed semantic summaries, predicted state deltas, and constraint signals through the Knowledge Plane.

This exchange forms an iterative semantic negotiation process through which agents align their world models, resolve resource conflicts, and converge toward mutually consistent control decisions. Semantic routing adaptation, load redistribution, and cooperative interference mitigation thus emerge as collective behaviors rather than centrally imposed commands.

When semantic negotiation fails to resolve persistent conflicts within a bounded number of iteration rounds due to conflicting world models, incompatible intents, or environmental ambiguity, Kraken activates hierarchical escalation mechanisms. Local edge-level agents escalate unresolved disputes to infrastructure-level agents that possess broader contextual awareness and longer planning horizons. If conflicts persist at this level, domain-level agents synthesize global coordination policies that reconcile local disagreements and ensure system-level stability.

\end{itemize}

These four components form a closed and self-reinforcing cognitive loop. Infrastructure signals are abstracted into semantic representations, embedded into structured memory, processed through generative planning mechanisms, and materialized as coordinated control actions. These actions reshape the physical environment, generating new observations that re-enter the perception stage. Through continuous iteration, agents adapt to non-stationarity, environmental uncertainty, and cross-domain coupling while preserving alignment with task-level objectives.

Collectively, the Agent Plane constitutes the distributed cognitive layer of Kraken. By structuring each network node around a perception–memory–planning–action architecture and deploying agents hierarchically across edge, infrastructure, and domain levels, Kraken enables scalable collective intelligence to emerge from coordinated multi-agent interaction. Nevertheless, autonomous cognition must remain anchored to system-wide intents and long-term objectives. This requirement motivates the final architectural layer, the Knowledge Plane, which provides global objective definitions, consistency validation, and overarching policy coherence.

\subsection{Kraken Knowledge Plane}
\begin{figure}[h]
\centering
\includegraphics[width=\linewidth]{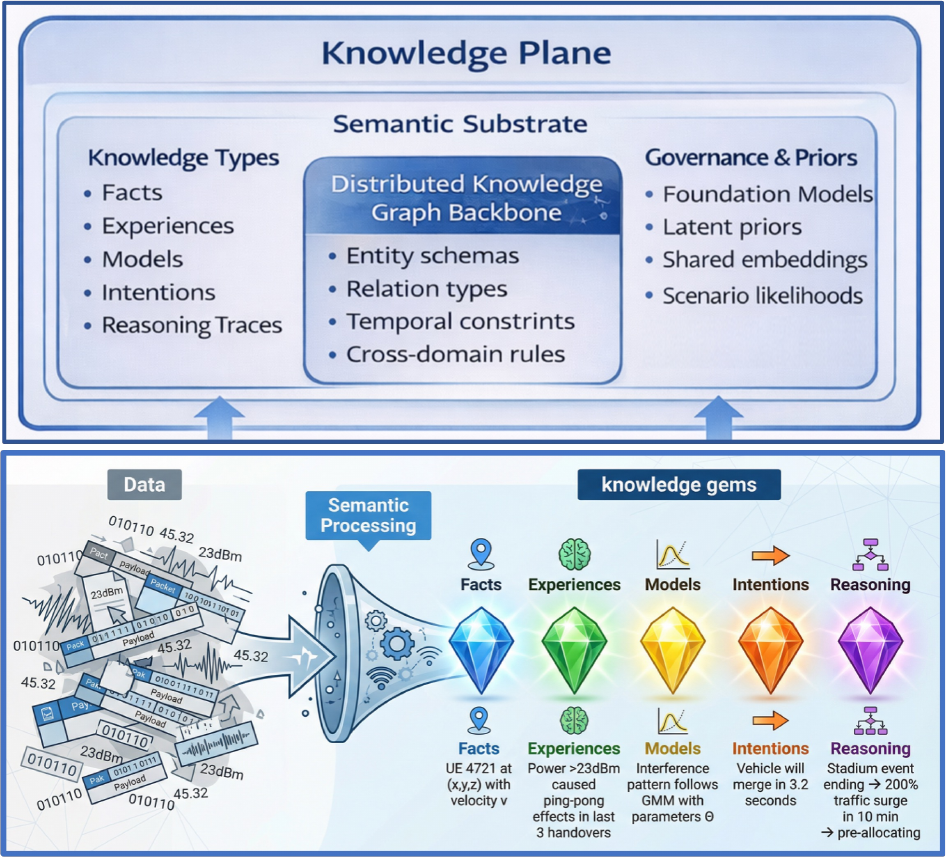}
\caption{Knowledge types in 6G. Knowledge objects include facts (with confidence and temporal validity), experiences (historical patterns), models (parametric representations), intentions (future objectives), and reasoning traces (explanations). Each type enables different forms of reasoning under uncertainty.}
\label{f4}
\end{figure}

While the Agent Plane enables distributed cognition at the operational level, sustainable intelligence in 6G systems requires a higher abstraction layer that defines, structures, validates, and governs knowledge itself. This role is fulfilled by the Knowledge Plane. Unlike conventional control planes that primarily disseminate configuration parameters, this layer maintains the semantic backbone of the network, defines system-wide intents, enforces consistency constraints, and ensures that distributed agents remain aligned with long-term objectives. Traditional networks exchange data in the form of raw measurements, protocol headers, and payload bits. Such data lacks operational structure beyond encoding and transport, leaving interpretation to higher-layer applications. In contrast, knowledge represents data enriched with contextual relationships, uncertainty estimates, and actionable semantics, as summarized in Fig.~\ref{f4}. In tightly coupled 6G ecosystems where communication, sensing, computation, and control continuously interact, knowledge becomes the primary abstraction exchanged among intelligent entities. Also, the distinction between data and knowledge is operationally significant. Exchanging structured knowledge objects rather than raw measurement streams reduces signaling overhead and enables coherent decision-making under partial observability, since agents operate on abstractions encoding causal and relational dependencies rather than isolated signals.

In this context, knowledge manifests in multiple structured forms, as illustrated in Fig.~\ref{f4}. First, it includes \emph{facts}, such as the location and velocity of a user equipment, annotated with confidence levels, temporal validity intervals, and provenance metadata. Second, it encompasses \emph{experiences}, which encode historically observed patterns that influence future coordination decisions. Third, knowledge includes \emph{models} that compress observations into predictive or generative representations, enabling abstraction beyond raw measurements. In addition, knowledge incorporates \emph{intentions}, representing declared or inferred future objectives such as mobility maneuvers, service-level targets, or resource constraints. Finally, structured \emph{reasoning traces} document the rationale behind coordination decisions under anticipated conditions, supporting interpretability and cross-agent transparency.

To support these knowledge types, the Knowledge Plane maintains a global semantic substrate built upon a distributed knowledge-graph backbone. This substrate defines entity schemas, relation types, temporal constraints, and cross-domain consistency rules that span the communication, computation, mobility, and service domains. While individual agents maintain local memory representations, the Knowledge Plane ensures cross-domain coherence by synchronizing ontologies, validating relational integrity, detecting semantic drift, and resolving representational conflicts. It therefore functions as a meta-memory and consistency-regulation layer rather than a centralized decision engine.

Complementing these structured symbolic representations, the Knowledge Plane integrates foundation models that capture large-scale statistical priors across heterogeneous network domains. These models are trained on aggregated observations such as traffic patterns, mobility traces, interference dynamics, and service behaviors, enabling them to learn transferable latent structures beyond locally observable measurements. Importantly, foundation models do not participate in real-time control loops. Instead, they function as epistemic prior generators that provide statistical guidance for generative planning within the Agent Plane while preserving decentralized operational control.

The interaction between foundation models and the distributed knowledge graph follows a dual-memory architecture that separates explicit dynamic knowledge from implicit statistical priors. The knowledge graph maintains explicit, structured, real-time knowledge, such as the current network topology, active sessions, measurement streams, and mobility states, as entities enriched with temporal validity intervals, confidence scores, and provenance metadata. In contrast, foundation models capture implicit statistical knowledge learned from historical data, including typical traffic distributions, interference correlations, mobility behaviors, and cross-domain service patterns.

These two memory systems interact through a learned embedding projection interface. Knowledge graph entities are mapped into the latent representation space of the foundation model using lightweight graph neural networks, allowing the model to condition its statistical priors on current context through cross-attention mechanisms. When queried, foundation models generate predictions or policy suggestions referencing specific graph entities. The outputs are subsequently grounded back into the knowledge graph as derived knowledge objects annotated with explicit provenance and confidence estimates. This interaction establishes a feedback loop in which the knowledge graph provides real-time grounding that constrains potential hallucinations, while foundation models enrich the graph with inferred relationships and predictive annotations.

To avoid real-time bottlenecks, synchronization between the two layers relies on temporal decoupling and hierarchical inference. Foundation models operate asynchronously and are typically updated through offline retraining on aggregated telemetry at hourly or daily intervals. Agents follow a two-tier inference strategy. In the fast path, agents query the knowledge graph directly to obtain current facts and relational context, enabling microsecond-level responses suitable for real-time coordination. In the slow path, foundation models are invoked only when graph information alone is insufficient, for example during high uncertainty, rare event prediction, or novel intent synthesis, where millisecond-level latency remains acceptable.

The interaction interface is exposed through a standardized Knowledge Plane API. Agents submit compressed semantic contexts, such as latent state representations or intent descriptors, and receive structured responses, including predicted state distributions, similarity scores to known operational patterns, or embeddings to support cross-domain transfer. Queries are asynchronous and cached at the edge to maintain real-time performance, while frequently requested inferences can be served through distilled lightweight models deployed closer to edge nodes.

Operational robustness is ensured through strict validation and versioning mechanisms. Foundation model updates are performed offline using aggregated and anonymized telemetry and are validated within the Network Digital Twin prior to deployment. All model outputs are checked against current knowledge graph facts before acceptance into the Knowledge Plane. If predictions conflict with observed measurements, for instance when a mobility prediction contradicts measured user velocity, the output is rejected and logged for retraining. Agents may also request rollback to previous model versions if monitoring detects semantic drift or degraded predictive performance.

Beyond representation and statistical priors, the Knowledge Plane defines system-wide goals and governance mechanisms. It maintains an intent repository that translates operator-defined service objectives into formalized task-level constraints and utility structures. Service-level agreements, safety requirements, energy budgets, fairness policies, and regulatory constraints are encoded into structured templates that parameterize the Lagrangian coordination objectives executed by individual agents.

Periodic global validation processes evaluate aggregate semantic distortion levels, task success distributions, resource-utilization fairness, and stability indicators across domains. When misalignment is detected, the Knowledge Plane adjusts constraint multipliers, reweights intent priorities, or refines semantic schemas, thereby preventing long-term drift between local adaptation and global objectives.

Through this governance loop, the Knowledge Plane performs meta-level feedback control over distributed intelligence. It does not micromanage operational actions. Instead, it shapes the objective landscape within which agents coordinate. Agents perceive, remember, plan, and act locally, while the Knowledge Plane defines how knowledge is structured, how consistency is enforced, and which objectives ultimately guide coordination.

This hierarchical yet autonomous organization ensures that latency-critical decisions remain local, while context-intensive strategic alignment is maintained at higher abstraction levels. By combining structured knowledge representations, statistical priors, and global intent governance, Kraken evolves from a collection of adaptive nodes into a coherent and self-consistent collective intelligence architecture.

\section{Enabling Technologies and Methodologies}
Realizing the Kraken vision requires more than architectural innovation. It requires operational mechanisms that ensure safe deployment, continuous validation, scalable learning, and lifecycle governance. While the previous sections introduced the transition toward knowledge-centric collective intelligence, this section presents the enabling technologies that operationalize that transformation. The discussion proceeds across three complementary layers of enablement. First, Open Radio Access Network (O-RAN) provides the standardized execution substrate in which distributed generative agents operate. Second, Machine Learning Operations (MLOps) governs the lifecycle of these agents at scale. Third, Network Digital Twins establish a simulation-to-reality bridge, safeguarding deployment through structured validation. Together, these mechanisms transform Kraken from a conceptual architecture into a deployable 6G intelligence framework.

\begin{figure*}[h]	
    \centering
    \includegraphics[width=\linewidth]{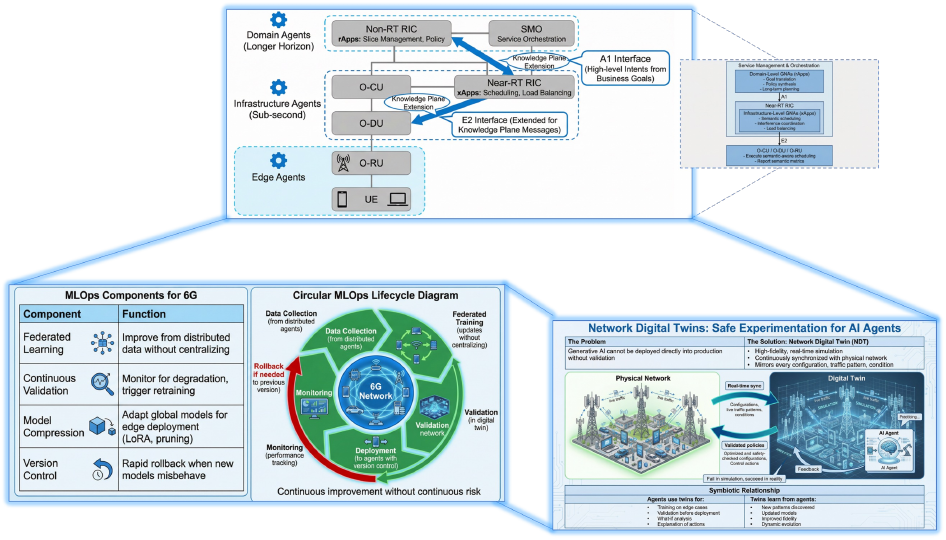}
	\caption{Operational enablement of KRAKEN: integration with Open Radio Access Network (O-RAN) control hierarchy, lifecycle governance through Machine Learning Operations (MLOps), and validation via Network Digital Twins forming a closed-loop intelligence framework.}
	\label{fig:enabling}
\end{figure*}

\subsection{Integration with O-RAN}

Open and disaggregated radio architectures provide a practical deployment substrate for Kraken systems. The Open Radio Access Network (O-RAN) framework defines a hierarchical control architecture composed of the Service Management and Orchestration (SMO) layer, the Non-Real-Time RAN Intelligent Controller (Non-RT RIC), and the Near-Real-Time RAN Intelligent Controller (Near-RT RIC), interconnected through standardized interfaces, as illustrated in Fig.~\ref{fig:enabling}. This architecture naturally accommodates the multi-plane organization of Kraken and enables distributed intelligence across multiple temporal scales.

At the non-real-time layer, domain-level agents operate within the SMO and the Non-RT RIC. The SMO supervises service orchestration, lifecycle management, and global policy coordination across the radio access network. Within the Non-RT RIC, RAN applications (rApps) perform long-horizon reasoning tasks such as slice management, policy refinement, and intent translation. These agents interact closely with the Knowledge Plane, transforming operator-defined objectives into structured coordination constraints. High-level intents and policy abstractions are conveyed to lower layers through the A1 interface, ensuring that long-term service goals guide operational behavior.

At the near-real-time layer, infrastructure-level agents are implemented as extensible applications (xApps) within the Near-RT RIC. These agents execute latency-sensitive coordination functions including scheduling adaptation, load balancing, and interference mitigation. They obtain telemetry and state information from the O-RAN Centralized Unit (O-CU), which performs higher-layer protocol processing, and the O-RAN Distributed Unit (O-DU), which handles lower-layer real-time functions, via the E2 interface. Using generative reasoning mechanisms embedded in the Agent Plane, xApps transform infrastructure feedback into coordinated control decisions under strict timing constraints. In this configuration, the Near-RT RIC serves as the execution locus for sub-second collective intelligence.

The Knowledge Plane spans both layers of the O-RAN control hierarchy. At the non-real-time level, it maintains global goal structures, statistical priors, and policy abstractions. At the near-real-time level, it supports semantic summaries of network state and structured performance indicators derived from infrastructure telemetry. High-level intents propagate downward through the A1 interface, while semantically abstracted performance information propagates upward through structured reporting mechanisms. This bidirectional exchange establishes a closed alignment loop between service objectives and radio-level actions.

Below the RIC hierarchy, the O-RAN Radio Unit (O-RU) performs physical-layer transmission functions, while the O-CU and O-DU implement higher- and lower-layer protocols respectively. User equipment and edge-level entities generate environmental observations and traffic dynamics, which are abstracted before being integrated into the Knowledge Plane. Consequently, higher-layer reasoning operates on structured semantic representations rather than raw radio signals.

Through this integration, O-RAN supplies the standardized operational backbone while Kraken embeds a distributed cognitive layer within that hierarchy. The architecture therefore preserves interface compliance and modularity while enabling knowledge-driven, goal-aligned coordination across heterogeneous 6G domains.

\subsection{MLOps for 6G: Lifecycle Governance of Distributed Intelligence}
While O-RAN defines where generative agents execute, Machine Learning Operations (MLOps) determines how these agents evolve, adapt, and remain trustworthy at scale. Kraken envisions large populations of distributed agents operating across heterogeneous infrastructure components and edge devices. Managing such an ecosystem requires systematic governance over training, validation, deployment, monitoring, updating, and retirement throughout the model lifecycle.

At 6G scale, thousands of models may operate across millions of devices, making manual oversight infeasible. MLOps extends DevOps principles to artificial intelligence systems by introducing automation, traceability, validation pipelines, and governance mechanisms across the entire model lifecycle, as illustrated in Fig.~\ref{fig:enabling}. In 6G environments, this lifecycle must additionally address decentralized data ownership, privacy preservation, hardware heterogeneity, and highly non-stationary network dynamics. The lifecycle begins with distributed data collection from edge and infrastructure agents. Rather than centralizing raw data, federated learning enables collaborative model improvement through decentralized parameter aggregation. Edge agents train local models using locally observed traffic patterns, mobility dynamics, and radio conditions, and transmit parameter updates to infrastructure-level aggregators. These aggregators refine global parameters and redistribute them to participating nodes, enabling collective improvement of shared generative priors while preserving data locality and privacy.

Before activation in production environments, updated models undergo structured validation procedures, frequently executed within Network Digital Twins. Validation evaluates coordination stability, goal alignment, semantic distortion behavior, and robustness under simulated edge cases and rare operational scenarios. Only models that satisfy predefined acceptance thresholds are promoted through version-controlled deployment pipelines. Once deployed, continuous monitoring tracks coordination metrics, stability indicators, and distortion-related measures in real time. When degradation is detected due to distributional drift or environmental changes, automated pipelines initiate retraining, specialization, or rollback procedures. Model compression techniques, including parameter-efficient fine-tuning, low-rank adaptation, and structured pruning, adapt generative models to resource-constrained devices while preserving reasoning capability and operational efficiency.

Within Kraken, MLOps functions as the governance layer of distributed cognition. It closes the loop among observation, adaptation, and controlled evolution, ensuring that collective intelligence continues to improve without compromising reliability, safety, or operational stability.

\subsection{Network Digital Twins: The Simulation-to-Reality Bridge}

Generative coordination mechanisms should not be deployed directly into operational 6G networks without structured validation. Network Digital Twins provide a safeguarded intermediate environment that bridges experimentation and live deployment. A Network Digital Twin is a high-fidelity, continuously synchronized virtual representation of the physical radio access network. Configuration states, traffic dynamics, topology evolution, interference patterns, and mobility behaviors are mirrored with bounded synchronization delay. This synchronization enables agents to evaluate coordination strategies in an operationally representative yet risk-isolated environment.

Within Kraken, interaction with the twin is tightly integrated into the MLOps lifecycle. During development and retraining phases, agents generate synthetic experience within the twin, including rare events and extreme operational scenarios. Candidate coordination policies are evaluated against safety constraints, stability conditions, and goal-consistency rules defined by the Knowledge Plane. Counterfactual analyses further quantify the effects of alternative decisions on service-level outcomes across varying network states. Following deployment, semantically abstracted telemetry from operational agents is fed back into the twin to refine predictive models. Observed congestion dynamics, mobility shifts, and interference patterns continuously recalibrate the virtual environment, allowing the twin to evolve alongside the physical network. Over time, this process transforms the twin into a predictive mirror capable of anticipating system behavior under emerging conditions. By decoupling experimentation from live execution while maintaining structural synchronization, Network Digital Twins enable safe innovation in large-scale network intelligence systems. In combination with O-RAN integration and MLOps governance, they complete Kraken’s closed-loop intelligence stack: standardized execution, lifecycle regulation, and validated evolution across emerging 6G networks.

\section{Case Studies: Generative AI in Action}
To ground the Kraken architecture in operational reality, we present three representative case studies. Each case follows a common structure: (i) system scenario, (ii) limitations of conventional approaches, (iii) Kraken-based realization across the Infrastructure, Agent, and Knowledge planes, and (iv) quantitative feasibility indicators. The numerical values reported in these examples are derived from representative results in the literature on semantic communication, predictive rendering, and distributed sensing systems, and are used to illustrate the potential efficiency gains achievable when such capabilities are integrated within the Kraken framework rather than representing measurements from a complete implementation. Together, the case studies illustrate how generative reasoning, semantic communication, and goal-oriented coordination enable capabilities beyond conventional 6G designs.

\subsection{Case Study 1: Cooperative Autonomous Driving}

\textit{Scenario:} 
A fleet of autonomous vehicles approaches an urban intersection characterized by partial GPS occlusion, multipath propagation, and fluctuating cellular coverage. Vehicles must coordinate merging, yielding, and crossing maneuvers under sub-second latency constraints while maintaining collision-free operation.

\textit{Conventional Approach:} 
In conventional architectures, vehicles stream raw sensor data such as camera feeds and LiDAR point clouds to a roadside unit or edge server. The infrastructure reconstructs a global scene representation and transmits maneuver decisions back to vehicles. This design consumes substantial uplink bandwidth, introduces round-trip latency, and becomes fragile under congestion or intermittent connectivity.

\textit{Kraken Realization:} 
Within the Agent Plane, each vehicle hosts an edge-level Generative Network Agent that performs local perception and semantic abstraction. Rather than transmitting raw sensor streams, vehicles exchange structured knowledge objects including occupancy grids, predicted trajectories with confidence intervals, and explicit intent declarations such as estimated crossing times. At the Infrastructure Plane, a roadside agent maintains a generative world model of the intersection. Incoming semantic descriptors are fused to detect potential trajectory conflicts and evaluate alternative coordination strategies. Generative planning explores candidate policies including temporal spacing adjustments, speed modulation, and dynamic priority reassignment. Coordinated intent messages are then disseminated to participating vehicles.  At the Knowledge Plane, statistical priors on intersection geometry, historical traffic patterns, and regulatory safety constraints guide the optimization process. Coordination decisions minimize a joint objective combining collision risk, maneuver completion time, and energy consumption. Vehicles execute locally optimized actions while remaining aligned with system-wide objectives through iterative semantic negotiation.

\begin{table*}[t]
	\centering
	\caption{Comparison of Kraken-enabled coordination across representative 6G applications.}
	\label{tab:case_comparison}
	\scriptsize
	\renewcommand{\arraystretch}{1.3}
	
	\begin{tabular}{p{3cm} p{3.2cm} p{3cm} p{3.2cm} p{2.3cm}}
		
		\toprule
		
		\textbf{Application} &
		\textbf{Semantic Abstraction (Agent Plane)} &
		\textbf{Predictive Reasoning (Infrastructure Plane)} &
		\textbf{Knowledge Coordination (Knowledge Plane)} &
		\textbf{Key Benefit} \\
		
		\midrule
		
		Autonomous Driving
		& Occupancy grids, predicted trajectories, intent messages
		& Intersection world model for conflict prediction
		& Safety constraints and traffic priors
		& 70--85\% uplink reduction, $\leq 100$ ms coordination \\
		
		XR Distributed Rendering
		& Scene descriptors, gaze direction, interaction intent
		& Predictive rendering using gaze trajectories
		& Persistent digital twin of factory environment
		& 10--20$\times$ bandwidth reduction, 15--25 ms latency \\
		
		Distributed Acoustic Sensing
		& Event signatures, anomaly scores, signal summaries
		& Structural response model for anomaly detection
		& Bridge digital twin for long-term risk analysis
		& Up to 100:1 sensing compression \\
		
		\bottomrule
		
	\end{tabular}
\end{table*}

Importance-aware source coding results indicate that semantic compression can reduce uplink bandwidth by approximately 70--85\% relative to raw sensor streaming while preserving task-relevant information. Multi-agent simulations using CARLA and SUMO further indicate that coordination converges within three to five negotiation rounds, introducing approximately 40--60 ms of additional latency. This delay remains well within the typical 100 ms cooperative maneuver window required for safe autonomous intersection management \cite{JI2024109751}, \cite{s21113783}.

\subsection{Case Study 2: Immersive Extended Reality with Distributed Rendering}

\textit{Scenario:} 
Lightweight XR headsets enable remote maintenance operations on a factory floor that includes static machinery, moving personnel, and dynamic digital overlays. Realistic rendering requires low motion-to-photon latency and stable perceptual quality under wireless resource constraints.

\textit{Conventional Approach:} 
In conventional architectures, headsets transmit raw video frames to an edge server for object detection, scene understanding, and rendering. The rendered frames are then encoded and streamed back to the device, forming a latency-sensitive feedback loop that is highly vulnerable to bandwidth fluctuations and network congestion.

\begin{figure}[h]	
    \centering
    \includegraphics[width=\linewidth]{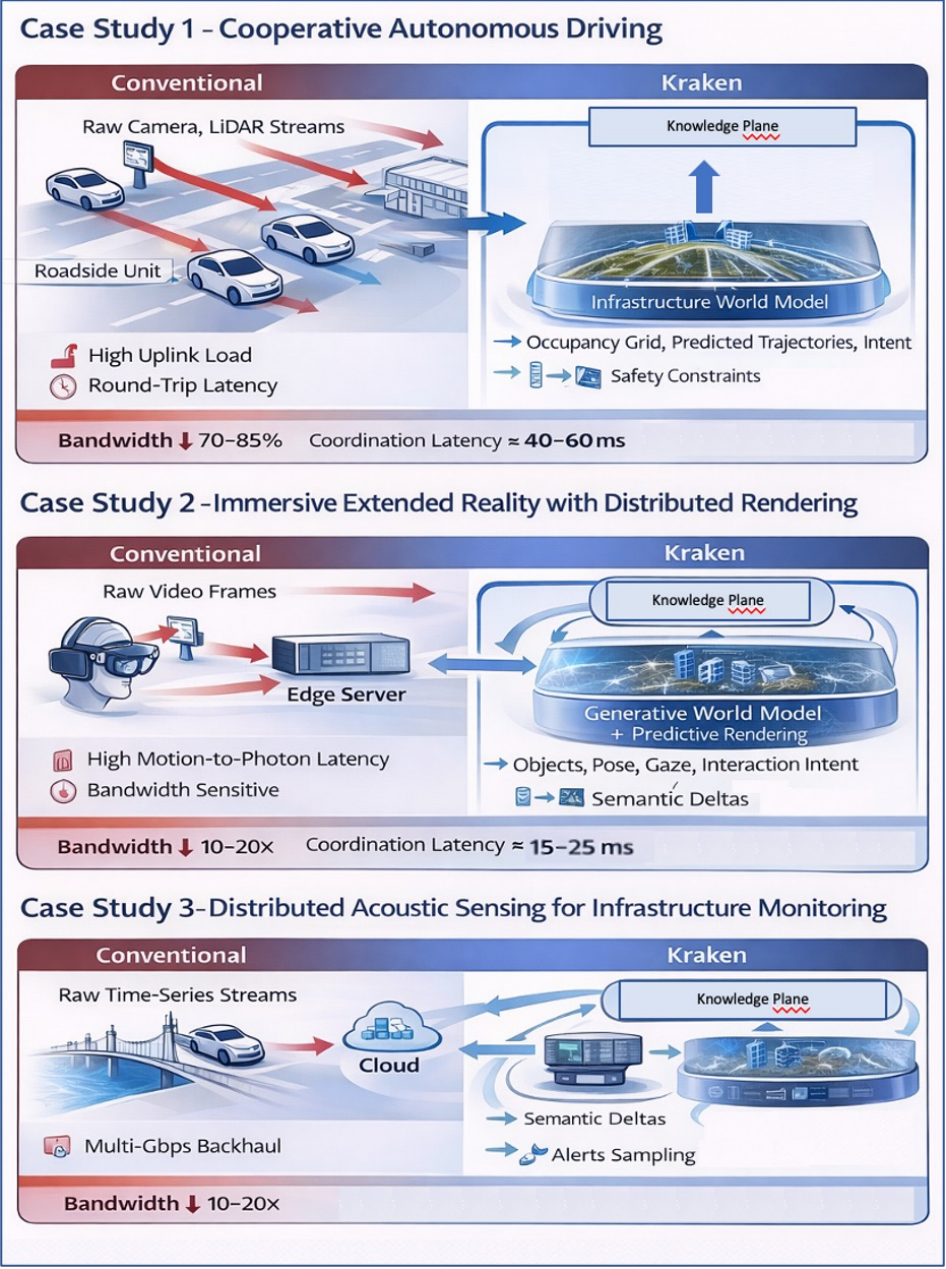}
	\caption{Representative KRAKEN case studies illustrating semantic abstraction at the edge, generative coordination at the infrastructure, and knowledge-plane alignment across autonomous driving, XR rendering, and distributed sensing scenarios.}
	\label{cases}
\end{figure} 

\textit{Kraken Realization:} 
Within the Agent Plane, the headset performs semantic scene extraction, generating compact descriptors including detected objects, user pose, gaze direction, and interaction intent. Instead of streaming full video frames, only these semantic descriptors are transmitted to the network. At the Infrastructure Plane, an edge-level generative world model synthesizes visual scenes by combining incoming semantic descriptors with cached environmental maps and statistical priors. Predictive rendering anticipates future view frustums from gaze trajectories, enabling rendering tasks to be partially precomputed. When predictions remain accurate, only semantic deltas representing state changes or intent updates are exchanged between the headset and the edge infrastructure. At the Knowledge Plane, a persistent digital twin of the factory environment maintains structural maps, calibration parameters, and long-term environmental priors. These knowledge objects support consistent scene reconstruction across devices and sessions while guiding coordination policies. Optimization objectives jointly consider perceptual fidelity, motion-to-photon latency, and headset energy consumption.

The semantic video analytics literature reports bandwidth reductions of approximately 10--20$\times$ for perception-driven tasks with minimal loss of accuracy. Predictive and gaze-contingent rendering techniques further reduce motion-to-photon latency from approximately 50--80~ms to 15--25~ms \cite{10.1145/3306307.3328201}, \cite{inbook}. These values align with the stringent latency and perceptual consistency requirements envisioned for 6G XR services.

\subsection{Case Study 3: Distributed Acoustic Sensing for Infrastructure Monitoring}

\textit{Scenario:} 
A smart city deploys fiber-optic distributed acoustic sensing along a bridge for structural health monitoring and traffic characterization. Continuous high-volume time-series data must be analyzed to detect anomalies such as crack formation, abnormal vibration patterns, and load classification events.

\textit{Conventional Approach:} 
Raw sensing traces are streamed to centralized cloud servers for analysis. This approach imposes substantial backhaul demand and introduces latency in anomaly detection. While edge processing can reduce data volume, many existing solutions rely on static, manually engineered features that offer limited adaptability to changing environmental conditions.

\textit{Kraken Realization:} 
Within the Agent Plane, sensing nodes host Generative Network Agents that perform semantic abstraction directly at the source. Rather than transmitting raw waveforms, agents extract structured knowledge objects including event timestamps, anomaly scores, signal signatures, and classification confidence levels. At the Infrastructure Plane, a generative world model trained on historical sensing data captures the normal distributions of structural response under varying environmental conditions. Deviations between predicted and observed signals trigger anomaly alerts accompanied by probabilistic explanations. This predictive reasoning enables early detection of abnormal structural behavior. At the Knowledge Plane, observations from multiple sensing agents are correlated across spatial locations to identify propagating vibration modes and resonance shifts. A continuously updated digital twin of the bridge supports predictive maintenance scheduling, long-term risk assessment, and coordinated infrastructure monitoring.

Industrial IoT studies report semantic compression ratios up to 100:1 while maintaining high anomaly-detection accuracy \cite{10825287}, \cite{unknownxx}. Such reductions can lower backhaul requirements from multi-Gbps sensing streams to hundreds of Mbps without sacrificing structural intelligence.

Across all three case studies, a consistent architectural pattern emerges, as summarized in Table~\ref{tab:case_comparison}. Raw signals are abstracted into structured semantic representations at the edge. Generative world models enable predictive reasoning and anomaly interpretation at the infrastructure level, while the Knowledge Plane enforces global consistency and long-horizon coordination across distributed agents. This layered transformation decouples bandwidth-intensive data transport from coordination logic, enabling scalable intelligence without overwhelming network resources. Decision-making therefore becomes both context-aware and resource-efficient across heterogeneous deployment scenarios. These examples illustrate that Kraken is not an application-specific solution but a reusable coordination paradigm for distributed collective intelligence in 6G systems. By explicitly separating semantic abstraction, generative reasoning, and goal alignment across architectural planes, Kraken reduces communication overhead, improves robustness under uncertainty, and maintains coherence between local autonomy and global system objectives.


\begin{table*}[t]
\centering
\caption{Engineering deployment dimensions, enabling mechanisms, and evaluation signals for transitioning the KRAKEN architecture toward practical 6G systems.}
\label{tab:kraken_deployment}
\footnotesize
\renewcommand{\arraystretch}{1.05}
\setlength{\tabcolsep}{6pt}

\begin{tabular}{p{4cm} p{6cm} p{6cm}}
\toprule
\textbf{Deployment dimension} & \textbf{Required mechanisms} & \textbf{Evaluation signals} \\
\midrule

\textbf{Complexity--practicality balance} &
\begin{itemize}
\item Semantic compression
\item Selective semantic exchange
\item Bounded negotiation
\item Energy-aware model partitioning
\end{itemize}
&
\begin{itemize}
\item Semantic efficiency metrics
\item Control-loop latency decomposition
\item Negotiation convergence indicators
\end{itemize} \\

\midrule

\textbf{Computational feasibility} &
\begin{itemize}
\item Device-class-aware model scaling
\item Quantization, pruning, and distillation
\item Latency-aware inference scheduling
\end{itemize}
&
\begin{itemize}
\item Latency and power budget compliance
\item Accuracy–cost trade-off
\item Responsiveness under urgency
\end{itemize} \\

\midrule

\textbf{Scalability of coordination} &
\begin{itemize}
\item Hierarchical aggregation ($k \ll N$)
\item Event-triggered negotiation
\item Compact semantic objects
\end{itemize}
&
\begin{itemize}
\item Coordination overhead vs.\ $N$
\item Convergence time
\item Task performance vs.\ coordination cost
\end{itemize} \\

\midrule

\textbf{Telco-grade reliability} &
\begin{itemize}
\item Deterministic control boundaries
\item Verification layers
\item Constrained action spaces
\item Explainability mechanisms
\item Robustness hardening
\end{itemize}
&
\begin{itemize}
\item Worst-case latency
\item Safety violation rate
\item Verification outcomes
\item Robustness under distribution shift or attacks
\end{itemize} \\

\midrule

\textbf{Semantic security \& privacy} &
\begin{itemize}
\item Knowledge provenance authentication
\item Confidence and temporal validation
\item Semantic anomaly detection
\item Privacy-preserving knowledge sharing
\end{itemize}
&
\begin{itemize}
\item Attack detection rate
\item Integrity verification success
\item Privacy leakage bounds
\item Security overhead
\end{itemize} \\

\bottomrule
\end{tabular}
\end{table*}

\section{Critical Analysis: From Vision to Viable Deployment}

The Kraken architecture departs from conventional data-centric network designs by introducing a vertically integrated Knowledge Plane, distributed Generative Network Agents, and semantic negotiation mechanisms across hierarchical control layers. While this paradigm is essential for enabling goal-oriented autonomy in future 6G systems, it also introduces non-trivial operational, computational, and engineering challenges. A realistic path toward deployment therefore requires a structured examination of architectural overhead, scalability limits, computational feasibility across heterogeneous infrastructure layers, and the maturity of the underlying foundation models. The following subsections analyze these key dimensions.
\subsection{Complexity--Practicality Trade-off in Knowledge-Centric Coordination}

Kraken shifts network intelligence from reactive signal exchange toward proactive semantic reasoning. This transformation introduces new forms of overhead that must be carefully quantified to determine whether collective intelligence yields a net system benefit. The primary sources of architectural overhead include:

\begin{itemize}

\item {Signaling and Semantic Communication Overhead:}  
Agents exchange structured semantic objects including inferred states, confidence levels, intent declarations, and model synchronization updates. Without efficient semantic compression and selective exchange mechanisms, knowledge synchronization overhead could offset the communication savings achieved through semantic abstraction. In practice, Kraken avoids global synchronization of all knowledge objects. Instead, event-driven and hierarchical synchronization mechanisms propagate only semantic deltas associated with significant state changes or coordination events, while locally stable knowledge remains confined within edge clusters.

\item {Computational and Inference Latency:}  
Generative reasoning requires trajectory sampling, counterfactual evaluation, and constrained optimization. On resource-limited platforms such as edge devices and base stations, these operations may pose challenges to strict 6G latency requirements. In practical deployments, generative reasoning is hierarchically distributed across the network: lightweight distilled models execute short-horizon perception and prediction at edge devices, whereas deeper generative planning and scenario evaluation are performed by infrastructure-level agents with greater computational capacity.

\item {Synchronization and Consistency Costs:}  
Distributed world models require alignment across multiple agents. Iterative semantic delta exchange and belief reconciliation procedures may delay convergence if not bounded by stability constraints and negotiation limits. Importantly, the semantic objects exchanged during coordination are typically orders of magnitude smaller than the raw sensing or multimedia streams generated by applications, ensuring that the bandwidth consumed by knowledge synchronization remains substantially lower than the communication savings achieved through semantic abstraction.

\item {Knowledge Acquisition and Maintenance:}  
Updating large foundation models remains computationally intensive and may require significant energy and backhaul bandwidth, particularly in infrastructure-constrained deployments.

\end{itemize}

To ensure deployment viability, qualitative architectural benefits must be translated into measurable system-level criteria:

\begin{itemize}

\item {Semantic Efficiency Metrics:}  
Task-relevant information delivered per transmitted bit, accounting for negotiation overhead and semantic metadata exchange.

\item {Latency Decomposition of Generative Control Loops:}  
Breakdown of perception, memory retrieval, planning, and execution stages to determine the feasible depth of generative reasoning under real-time constraints.

\item {Convergence Analysis of Semantic Negotiation:}  
Modeling multi-agent alignment as a distributed consensus process within a semantic state space.

\item {Energy-Aware Model Partitioning:}  
Hierarchical placement of reasoning tasks across device, edge, and infrastructure layers to balance latency, energy consumption, and computational load.

\end{itemize}

\subsection{Computational Constraints and Optimization Strategies}

Deployment feasibility ultimately depends on aligning generative intelligence with the heterogeneous capabilities of the underlying infrastructure. Compute capacity, memory availability, power budgets, and latency constraints vary significantly across network layers. A uniform model deployment strategy is therefore impractical. Instead, Kraken adopts hierarchical specialization and adaptive execution strategies:

\begin{itemize}

\item {Device-Class-Aware Model Scaling:}  
User equipment and sensing devices require ultra-compact agents based on aggressive quantization, structured pruning, and distilled student models. Edge nodes can host moderately compressed models using mixed precision and low-rank adaptation. Regional and core infrastructure, by contrast, support near-full-scale foundation models employing sparse attention and conditional computation mechanisms. This hierarchical scaling ensures that latency-critical operations remain feasible on resource-constrained devices while computationally intensive reasoning is executed at infrastructure layers with sufficient processing capacity.

\item {Model Compression Techniques:}  
A combination of compression methods is required to reduce computational and memory footprints while preserving reasoning capability. These include integer quantization, structured and unstructured pruning, knowledge distillation, low-rank adaptation, sparse attention mechanisms, and mixture-of-experts architectures that activate subnetworks conditionally.

\item {Latency-Aware Inference Scheduling:}  
Agents dynamically adjust the depth of reasoning according to task urgency, uncertainty levels, and available computational resources. Shallow inference enables rapid responses under strict delay constraints, whereas deeper generative reasoning is invoked when uncertainty dominates or when coordination complexity increases.

\item {Adaptive Planning--Responsiveness Balance:}  
Dynamic control of trajectory exploration and planning horizon provides a principled mechanism for balancing decision quality with real-time responsiveness. By adapting the planning depth to the operational context, Kraken maintains responsiveness while preserving the benefits of generative reasoning.

\end{itemize}

\subsection{Scalability of Distributed Collective Intelligence}
Naive pairwise synchronization among $N$ agents leads to quadratic communication complexity $\mathcal{O}(N^2)$, which becomes infeasible at the scale envisioned for dense 6G deployments. Kraken addresses this limitation through hierarchical and selective coordination mechanisms that reduce coordination overhead while preserving collective intelligence.

\begin{itemize}

\item {Hierarchical Aggregation:}  
Agents are organized into clusters at edge and infrastructure layers. Local coordination occurs within clusters of size $k \ll N$, while higher-level agents mediate inter-cluster alignment. This hierarchical structure limits global synchronization overhead and enables scalable coordination across large deployments.

\item {Selective Interaction Policies:}  
Event-triggered semantic negotiation restricts coordination exchanges to situations involving significant model divergence, elevated task risk, or high uncertainty. By avoiding continuous synchronization, the system reduces unnecessary signaling overhead.

\item {Semantic Object Compression:}  
Exchanged knowledge objects such as intent declarations, predicted trajectories, and semantic state summaries remain compact relative to raw telemetry streams. This representation significantly reduces communication load during coordination.

\item {Sub-Quadratic Coordination Growth:}  
The combination of hierarchical clustering and selective negotiation limits the growth of coordination overhead as network scale increases. As a result, coordination complexity scales significantly more slowly than $\mathcal{O}(N^2)$, enabling practical deployment in ultra-dense network environments.

\end{itemize}
\subsection{From Foundation Models to Telco-Grade Reliability}

While foundation models offer powerful generative reasoning capabilities, their direct integration into telecommunications systems requires careful engineering to satisfy deterministic and safety-critical operational requirements. Unlike conventional AI applications, telecom infrastructure demands strict guarantees of latency, reliability, and operational stability. As a result, foundation models must be adapted into \emph{telco-grade} components through additional architectural safeguards and control mechanisms.

\begin{itemize}

\item {Predictable Latency and Deterministic Boundaries:}  
Generative models operate as high-level reasoning and orchestration components, while deterministic control modules handle strict real-time execution loops. This separation ensures that generative inference does not violate latency guarantees required by time-sensitive network functions.

\item {Constrained Action Spaces and Verification Layers:}  
Model outputs are restricted to predefined action spaces and validated through rule-based or formally verified control layers prior to execution. Such verification mechanisms prevent unsafe or inconsistent configurations from propagating into the operational network. For instance, before a Generative Network Agent’s suggested power adjustment, scheduling modification, or handover command is forwarded to the O-RAN Near-RT RIC for execution, it can be validated against a safe operational envelope defined by 3GPP specifications and operator policies, ensuring that generative reasoning cannot produce physically infeasible or protocol-violating configurations.

\item {Explainability and Decision Provenance:}  
Operational deployments require traceable reasoning paths that enable post-decision auditing and fault diagnosis. Structured knowledge representations within the Knowledge Plane provide decision provenance and enable explainable coordination across distributed agents.

\item {Robustness and Security Hardening:}  
Adversarial robustness training, input validation, and output verification mechanisms are required to protect against malicious inputs, model manipulation, and unsafe system behavior. These safeguards ensure that generative reasoning remains reliable under adversarial or uncertain operating conditions.

\end{itemize}

\subsection{Security and Privacy Architecture}

Semantic coordination introduces new attack surfaces that extend beyond conventional protocol-level threats. Because coordination decisions depend on shared semantic representations and distributed world models, adversarial manipulation of knowledge objects may propagate through multiple agents and influence system-wide behavior. Security mechanisms must therefore operate directly within the Knowledge Plane and protect both semantic communication and generative coordination processes.

\begin{itemize}

\item {Threat Model:}  
Potential attacks include semantic injection targeting distributed world models, confidence manipulation to bias prioritization decisions, replay of outdated knowledge objects, model extraction attempts, and adversarial inputs designed to induce unsafe generative behavior. Such attacks may disrupt coordination policies or degrade collective decision-making across agents.

\item {Knowledge Provenance Authentication:}  
Cryptographic signatures and hierarchical chain-of-trust mechanisms ensure the authenticity and integrity of exchanged semantic objects. Each knowledge object is associated with verifiable provenance metadata that allows agents to validate the origin and trustworthiness of shared information.

\item {Confidence and Temporal Validation:}  
Cross-agent verification and temporal consistency checks mitigate manipulation and replay attacks. Confidence intervals and validity windows attached to semantic objects allow agents to detect outdated or inconsistent knowledge states during negotiation and synchronization processes.

\item {Privacy-Preserving Knowledge Sharing:}  
Privacy protection mechanisms ensure that distributed coordination does not expose sensitive data. Differential privacy constrains information leakage in aggregated statistics, secure multi-party computation enables collaborative inference without revealing raw inputs, and homomorphic encryption preserves confidentiality during distributed model updates.

\end{itemize}

\subsection{Toward Empirical Validation}

The Kraken architecture presented in this article establishes a theoretical and architectural foundation for knowledge-centric 6G intelligence. Its design integrates semantic communication, generative reasoning, and goal-oriented coordination into a unified multi-plane framework. While the individual components underlying Kraken build upon established advances in semantic communication, multi-agent coordination, and generative modeling, the emergent system-level properties of the architecture, such as collective intelligence, distributed goal alignment, and generative coordination, require comprehensive empirical evaluation.

The quantitative indicators reported in the case studies are therefore interpreted as feasibility references derived from component-level results available in the literature. These results demonstrate the practicality of key enabling mechanisms, including semantic compression, predictive rendering, and distributed sensing analytics. However, they do not yet represent the end-to-end behavior of a fully integrated Kraken system operating across all architectural planes.

Several important research questions remain to be examined through system-level experimentation. For example, it remains necessary to evaluate whether distributed world-model alignment converges under realistic wireless channel dynamics and mobility patterns. Similarly, the relative advantages of generative agents compared to well-established baselines, such as conventional multi-agent reinforcement learning (MARL) approaches, must be assessed in terms of task success rate, coordination stability, and sample efficiency.

To address these questions, our ongoing research focuses on developing a full-stack system-level simulation environment that integrates semantic abstraction, generative planning, and the Knowledge Plane coordination mechanisms introduced in this paper. This simulator will support controlled experimental comparisons between Kraken and existing 5G and AI-native networking approaches across the representative scenarios presented in Section~VI. Such evaluations will enable systematic analysis of performance gains, coordination convergence behavior, and operational overhead, thereby providing quantitative evidence for the practical benefits and deployment feasibility of knowledge-centric collective intelligence in future 6G systems.

\subsection{Standardization and Ontology Governance}
A fundamental challenge for knowledge-centric networking is semantic interoperability across heterogeneous vendors, operators, and application domains. Unlike conventional protocol headers, knowledge objects carry structured semantic content, including contextual attributes, confidence levels, and temporal validity. Ensuring that these representations are interpreted consistently across distributed agents, therefore, requires explicit ontology governance and standardization mechanisms.

In the Kraken architecture, we envision a multi-stakeholder governance model for semantic definitions. A core upper ontology defining fundamental entities, relations, and temporal semantics is standardized by telecommunications bodies such as 3GPP and ITU. This upper ontology provides a common semantic foundation for representing network state, coordination intents, and knowledge objects across heterogeneous infrastructure components.

On top of this shared layer, domain-specific ontology extensions are defined collaboratively by vertical industry alliances and application ecosystems. For example, automotive consortia may define semantic representations for cooperative maneuvers and safety constraints, while industrial automation alliances may introduce schemas describing robotic operations, sensing events, or production workflows. These extensions remain compatible with the standardized upper ontology while enabling domain-specific expressiveness.

To maintain interoperability in dynamic environments, the Knowledge Plane supports runtime ontology mediation and negotiation mechanisms. When agents encounter semantic mismatches or previously unseen schema extensions, lightweight mediation protocols enable dynamic mapping between compatible concepts and attributes. This approach avoids the rigidity of a single static global schema while preserving interoperability across vendors and domains. Through this layered governance structure, Kraken balances standardization rigor with domain-specific flexibility. The resulting semantic ecosystem enables consistent interpretation of knowledge objects while supporting the diverse application landscapes expected in future 6G systems.

\subsection{Summary: From Conceptual Architecture to Validated System}

The Kraken architecture presented in this article establishes a rigorous conceptual foundation for knowledge-centric collective intelligence in future 6G systems. By integrating semantic communication, generative reasoning, and goal-oriented coordination across a unified multi-plane framework, it outlines how distributed intelligence can emerge from structured interaction among network entities. The engineering challenges associated with transitioning such an architecture toward practical deployment, including complexity management, computational feasibility, scalability, reliability, and security, were summarized earlier in Table~\ref{tab:kraken_deployment}.

Nevertheless, several aspects of the architecture require further refinement and empirical investigation. First, the algorithmic realization of the generative world model, including its neural architecture, training objectives, and planning algorithms, remains an open research direction that extends beyond the high-level architectural description provided in this work. Second, the emergent properties of the system, such as collective intelligence, semantic negotiation, and distributed goal alignment, must be quantitatively evaluated within controlled experimental environments.

To address these challenges, our immediate research effort focuses on developing a full-stack, event-driven simulation environment that instantiates the three Kraken planes. This simulator will enable systematic experimentation and quantitative comparison with existing 5G and AI-native baselines, including conventional multi-agent reinforcement learning (MARL) approaches, across the representative case-study scenarios presented in this article. Beyond proof-of-concept validation, the simulator will serve as an experimental platform for refining the algorithmic components underlying generative coordination and semantic reasoning.

By progressively bridging the gap between architectural vision and operational validation, this research program aims to transform Kraken from a conceptual blueprint into an empirically grounded foundation for knowledge-centric intelligence in future 6G networks.

\section{Transition Path: Evolutionary Deployment from 5G to Kraken-Enabled 6G}

\begin{figure*}[h]	
    \centering
    \includegraphics[width=\linewidth]{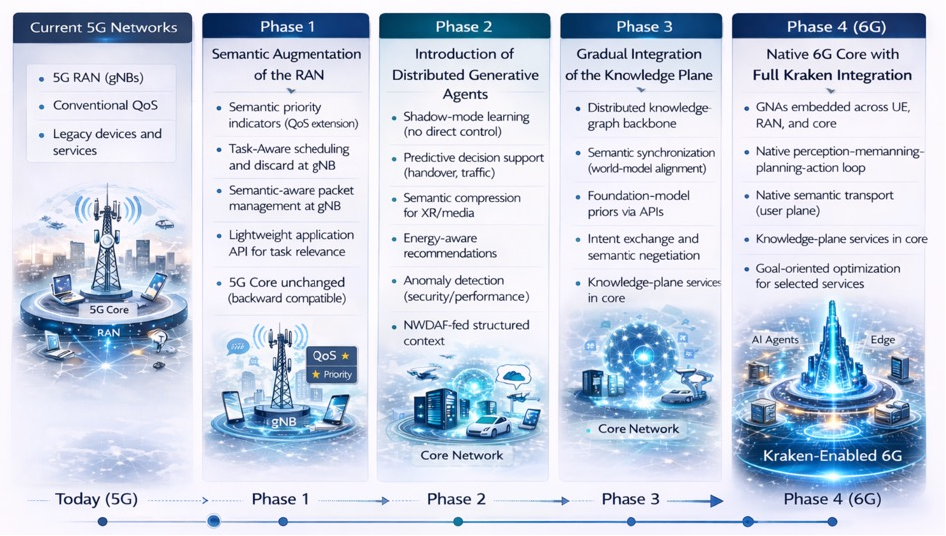}
\caption{Evolutionary transition from current 5G infrastructure to KRAKEN-enabled 6G collective intelligence. 
Phase~1 introduces semantic augmentation of the RAN through QoS priority flags and task-aware scheduling; 
Phase~2 deploys distributed Generative Network Agents (GNAs) at the edge in shadow mode; 
Phase~3 integrates a distributed knowledge plane enabling knowledge graphs and intent-based coordination; 
Phase~4 realizes a native 6G core with full KRAKEN integration and semantic communication capabilities.}
\label{evo}
\end{figure*}

As discussed in the previous section, the transition toward knowledge-centric networking introduces multiple engineering dimensions, including computational feasibility, coordination scalability, reliability, and security, summarized in Table~\ref{tab:kraken_deployment}. Realizing the Kraken architecture in practice, therefore, requires a staged deployment strategy that progressively addresses these dimensions within operational networks. Rather than replacing existing systems through a disruptive transition, the evolution toward Kraken-enabled networks must follow an incremental, standards-aligned pathway that gradually introduces semantic awareness, generative cognition, and knowledge-plane coordination while preserving backward compatibility. Fig.~\ref{evo} illustrates this evolutionary transition from contemporary 5G infrastructures toward native Kraken-enabled 6G operation.

\subsection{Phase 1: Semantic Augmentation of the RAN}

The transition begins by introducing semantic awareness into the existing 5G radio access network without modifying the core architecture. In this phase, applications provide lightweight semantic priority indicators that complement conventional QoS parameters. These indicators allow the gNB to differentiate traffic according to task relevance, enabling task-aware scheduling and packet discard decisions while maintaining the standard packet core.

Key developments in this phase include extending QoS class identifiers with semantic priority fields, implementing semantic-aware packet management at the gNB, and introducing lightweight application interfaces for declaring task relevance. Field trials focus on bandwidth reduction and robustness improvements in XR, IoT, and sensing workloads. Since semantic annotations remain optional, legacy devices continue to operate unchanged and receive conventional best-effort service. This phase demonstrates immediate value from semantic prioritization while requiring no changes to the 5G core.

\subsection{Phase 2: Introduction of Distributed Generative Agents}

The second phase introduces the first Generative Network Agents (GNAs), edge-resident intelligence modules co-located with user-plane functions and edge-cloud infrastructure. Initially, these agents operate in an observational mode, learning traffic dynamics and predicting network states without direct control authority. This shadow deployment ensures operational safety while enabling model training on real network telemetry.

As confidence grows, agents begin to provide bounded decision-support functions, such as predictive handover assistance, semantic compression of media streams, adaptive energy management for base stations, and anomaly detection in traffic patterns. During this stage, the 5G core remains the primary control plane, while GNAs function as an intelligent overlay. Existing analytics functions, including NWDAF, evolve to provide structured contextual information to agents, thereby enriching semantic perception without altering the network's control semantics. This phase demonstrates the operational feasibility of generative reasoning within live networks.

\subsection{Phase 3: Gradual Integration of the Knowledge Plane}

The third phase marks the emergence of a distributed knowledge-plane overlay interconnecting edge and infrastructure agents. A lightweight knowledge-graph backbone is deployed across edge and regional sites to enable semantic synchronization and shared world modeling. Foundation models are introduced as shared prior generators accessible through standardized interfaces, providing consistent semantic embeddings across domains.

Cross-agent coordination begins through intent exchange and semantic negotiation mechanisms, enabling limited goal-oriented optimization for selected mission-critical services such as industrial automation corridors and autonomous mobility zones. Legacy analytics components progressively evolve into knowledge-plane services that provide structured context rather than raw measurements. This phase requires targeted updates to core network functions while maintaining interoperability through protocol translation gateways and compatibility interfaces.

\subsection{Phase 4: Native 6G Core with Full Kraken Integration}

The final phase culminates in a native 6G architecture designed from the outset to align with the Kraken multi-plane model. Generative Network Agents become embedded within all network entities, including user equipment, access nodes, and core functions, enabling the perception–memory–planning–action cognitive loop to operate natively across protocol layers. Hardware acceleration at edge and core infrastructure supports real-time generative inference for mission-critical services.

Semantic transport becomes a first-class capability in the user plane, allowing knowledge objects and intent descriptors to flow alongside conventional data streams. Instead of exchanging only raw packets, the network can transmit structured semantic representations such as predicted trajectories, task descriptors, and context-aware state abstractions, reducing redundant transmission while preserving task-relevant information.

The Knowledge Plane functions as a global semantic substrate that interconnects distributed agents and maintains cross-domain consistency through shared ontologies, semantic schemas, and knowledge graphs. Foundation models provide shared statistical priors to support cross-domain reasoning, while network digital twins operate as native services that mirror operational states and enable the safe validation of coordination policies before deployment.

Despite this architectural transition, legacy 5G devices remain supported through compatibility interfaces exposed by the 6G infrastructure, ensuring service continuity during migration. This phased pathway de-risks deployment by aligning technological evolution with infrastructure lifecycles and standardization processes. Each stage delivers incremental operational value while preparing the substrate for subsequent capabilities. The transition path also clarifies research priorities across different time horizons: near-term work focuses on semantic augmentation of existing 5G systems and lightweight generative overlays, mid-term research addresses trustworthy distributed agents and knowledge-plane synchronization, and long-term efforts target native knowledge-centric 6G architectures with integrated semantic transport and collective intelligence. Through this staged evolution, Kraken capabilities emerge progressively while maintaining continuity with existing network ecosystems.

\begin{figure*}[h]	
    \centering
    \includegraphics[width=0.75\linewidth]{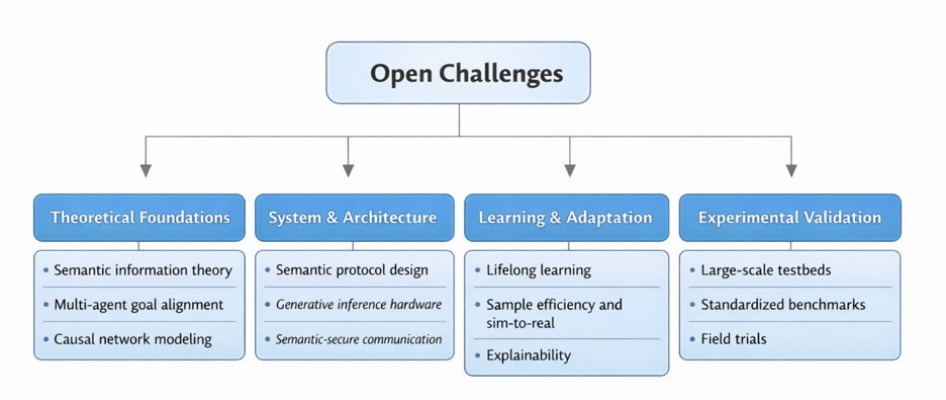}
\caption{Taxonomy of open challenges in KRAKEN, spanning theoretical foundations, system architecture, learning and adaptation, and large-scale experimental validation required for deployable knowledge-centric 6G intelligence.}
	\label{challenges}
\end{figure*}

\section{Open Challenges and Future Directions}

While Kraken establishes a coherent architectural framework for knowledge-centric collective intelligence in 6G systems, several fundamental research and engineering challenges must be addressed before large-scale deployment becomes feasible. These challenges span theoretical foundations, system architecture, learning dynamics, and experimental validation. The following subsections outline the most critical open research directions.

\subsection{Theoretical Foundations of Knowledge-Centric Networking}

The transition from data-centric communication toward semantic and goal-oriented coordination introduces theoretical questions that remain largely unresolved. Establishing rigorous foundations is therefore essential for the systematic analysis and design of knowledge-centric networking systems.

\begin{itemize}

\item {Semantic information theory:}  
Classical information theory quantifies limits for reliable bit transmission and reconstruction fidelity but does not capture task relevance or semantic distortion. A rigorous theory of semantic information is required to characterize rate–distortion trade-offs under task-oriented objectives. Such a framework must quantify the amount of semantic compression achievable for a given task utility and channel constraint, analogous to Shannon’s rate–distortion theory, but defined over knowledge representations rather than symbol sequences.

\item {Multi-agent goal alignment:}  
Kraken relies on distributed agents coordinating through semantic negotiation to satisfy global objectives. Although Lagrangian coordination provides a mathematical foundation, convergence properties under partial observability, delayed communication, heterogeneous objectives, and non-stationary environments remain largely unexplored. Establishing conditions for convergence toward globally optimal or bounded-suboptimal policies therefore represents a central theoretical challenge.

\item {Causal modeling of network dynamics:}  
Generative world models must capture causal dependencies rather than mere correlations to support counterfactual reasoning and proactive control. Integrating causal inference with deep generative modeling in networked systems, where interventions are costly and observations are partial, remains an open problem. Formalizing causal structure learning for communication–computation–mobility coupling is therefore essential.

\end{itemize}

\subsection{System and Architectural Realization Challenges}

Translating knowledge-centric coordination into operational infrastructure requires new protocol abstractions, hardware support mechanisms, and security primitives beyond conventional network design. The principal architectural challenges arise from embedding semantic exchange and generative inference into latency-constrained communication systems.

\begin{itemize}

\item {Semantic protocol design:}  
Knowledge-centric networking requires protocols capable of serializing structured semantic objects, negotiating ontologies among heterogeneous agents, and maintaining semantic consistency under bandwidth constraints. These protocols must also support dynamic schema evolution, confidence propagation, and graceful degradation. Standardization efforts within 3GPP and IETF will therefore need to define common semantic ontologies, exchange formats, and negotiation procedures.

\item {Hardware acceleration for generative inference:}  
Generative world models and reasoning agents rely on transformer-based architectures whose computational demands exceed those of conventional control algorithms. Achieving real-time performance at the edge requires specialized acceleration for sparse attention, quantized inference, and memory-efficient execution pipelines. Co-design of model architectures and base-station or edge system-on-chip platforms is therefore necessary to meet latency–power–capability trade-offs.

\item {Semantic-secure communication:}  
Knowledge exchange expands the attack surface beyond packet-level threats, enabling adversaries to inject false semantic objects, manipulate confidence annotations, or exploit reasoning chains. Security mechanisms must therefore operate directly in semantic space, including provenance authentication, cross-agent semantic validation, anomaly detection, and privacy-preserving knowledge sharing. Designing lightweight yet robust semantic security primitives remains an open architectural challenge.

\end{itemize}

\subsection{Learning and Adaptation in Non-Stationary 6G Environments}

Distributed generative agents must continuously adapt to evolving network conditions while preserving stability and reliability. Learning challenges arise from environmental drift, limited observations of rare events, and the need for interpretable decision processes.

\begin{itemize}

\item {Lifelong learning under environmental drift:}  
Ultra-dense 6G environments evolve due to traffic shifts, infrastructure upgrades, and changing user behavior. Generative agents must update their world models without catastrophic forgetting while preserving previously acquired knowledge. Mechanisms for detecting distributional shifts, triggering model recalibration, and maintaining stability in distributed continual learning remain largely underdeveloped.

\item {Sample efficiency and sim-to-real transfer:}  
Critical network events such as cascading failures or extreme congestion are inherently rare in observational data. Improving sample efficiency through meta-learning, transfer learning, and digital-twin-based simulation is therefore essential. Bridging the gap between simulated and operational network behavior remains challenging due to model mismatch and environmental stochasticity.

\item {Explainability and interpretability of generative decisions:}  
Autonomous knowledge-plane agents must provide reasoning traces that allow operators to understand why actions were selected, which evidence supported predictions, and which alternatives were considered. Extracting interpretable structure from deep generative models remains difficult. Neuro-symbolic integration combining structured knowledge graphs with neural representations offers a promising direction but still lacks scalable realizations.

\end{itemize}

\subsection{Experimental Validation and Benchmarking at Scale}

The viability of collective intelligence architectures cannot be established solely through simulation. Empirical validation across realistic deployments and standardized evaluation environments is required to assess both performance gains and operational overhead.

\begin{itemize}

\item {Large-scale experimental testbeds:}  
Realistic evaluation requires platforms capable of implementing semantic abstraction, distributed negotiation, and generative inference over operational wireless infrastructure. Such testbeds should expose measurable metrics including semantic compression ratios, negotiation convergence time, inference latency, and SLA compliance under dynamic traffic conditions. Establishing and maintaining these infrastructures represents a substantial community-level effort.

\item {Standardized benchmarks for semantic coordination:}  
Comparative evaluation of knowledge-centric networking approaches requires common task definitions, environmental dynamics, semantic performance metrics, and baseline implementations. Benchmarks must jointly measure goal-oriented performance and coordination overhead to capture the intelligence–complexity trade-off inherent in Kraken-like architectures.

\item {Long-term field trials in operational networks:}  
Extended deployments are necessary to reveal emergent behaviors, rare failure modes, and operational constraints that cannot be observed in controlled experiments. Demonstrating the viability of collective intelligence therefore requires sustained collaboration among academia, industry, and network operators.

\end{itemize}

\section{Conclusion}

This article has introduced Kraken, a unified architectural framework for realizing distributed collective intelligence in 6G wireless systems. Departing from conventional data-centric designs that optimize isolated protocol-level metrics, Kraken integrates three complementary pillars: semantic communication for meaning-aware representation, generative reasoning for structured adaptation under uncertainty, and goal-oriented optimization for explicit alignment with application-level intents. These capabilities are embedded within a multi-plane architecture comprising a semantic-aware Infrastructure Plane, distributed Generative Network Agents operating across edge and core, and a Knowledge Plane that maintains global semantic coherence and governs long-term objectives.

The contributions of this work are both conceptual and practical. Conceptually, Kraken reframes network intelligence as an emergent property of coordinated multi-agent interaction over structured knowledge representations rather than a centralized optimization or isolated AI function. In practice, the architecture is designed for incremental deployment from existing 5G infrastructure toward native 6G systems, with clear integration paths through O-RAN platforms, MLOps pipelines, and Network Digital Twins. The three case studies, cooperative autonomous driving, immersive XR rendering, and distributed acoustic sensing, illustrate how Kraken’s principles translate into measurable improvements across diverse 6G scenarios, with preliminary feasibility indicators supporting the practicality of semantic compression, generative planning, and goal-aligned coordination.

Kraken represents an architectural vision under active development. The authors are implementing its core components through a multi-year research program targeting a complete knowledge-centric network management framework. Initial prototypes focus on semantic-aware scheduling at the MAC layer, lightweight generative compression at the edge, and federated training of distributed world models across experimental testbeds. Subsequent work will report detailed algorithm design, large-scale experimental validation, and standardization proposals as the architecture matures. The phased transition pathway described in this article ensures that each incremental capability can deliver operational value while progressively building toward full Kraken realization.

Kraken is therefore positioned not as a static specification but as an evolving framework for knowledge-native 6G networks. By establishing architectural foundations, defining key abstractions, and demonstrating early feasibility, this work aims to guide future research toward networks that move beyond transporting bits to reasoning over meaning, coordinating distributed intelligence, and supporting complex cyber–physical systems. The transition from today’s data-centric infrastructures to tomorrow’s knowledge-centric collective intelligence will be gradual, but Kraken provides a coherent architectural direction for navigating that evolution.

	\bibliographystyle{IEEEtran}
	\bibliography{ref}

\begin{IEEEbiography}[{\includegraphics[width=1in,height=1.3in,clip,keepaspectratio]{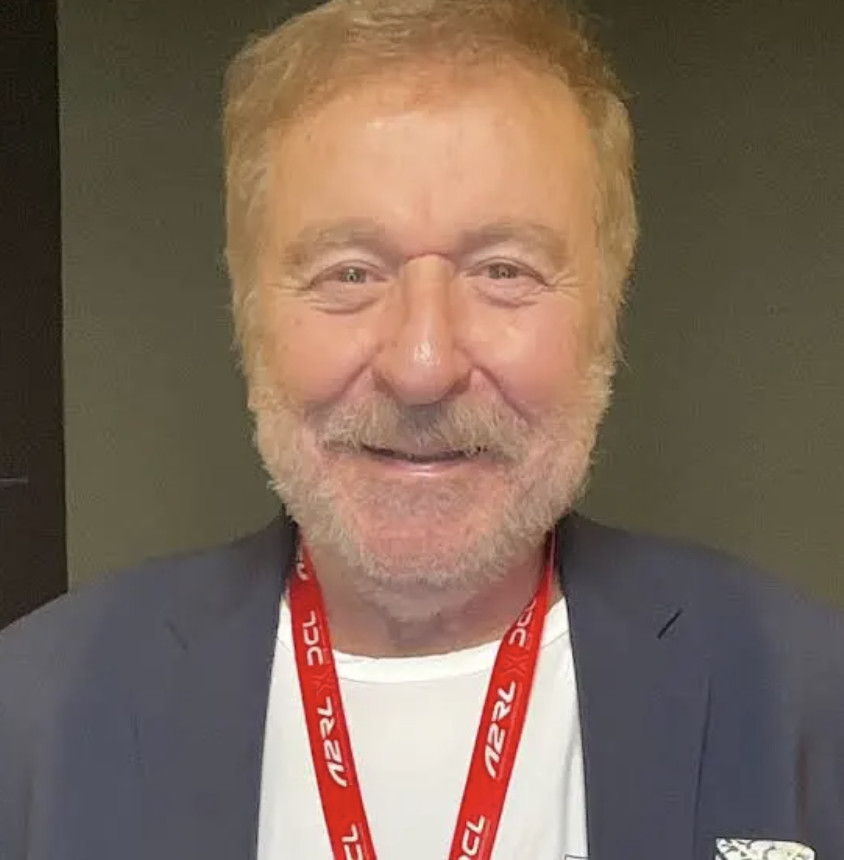}}]{Ian F. Akyildiz}, \textit{Life Fellow, IEEE}, received the B.S., M.S., and Ph.D. degrees in electrical and computer engineering from the University of Erlangen–Nürnberg, Germany, in 1978, 1981, and 1984, respectively. From 1985 to 2020, he held the prestigious Ken Byers Chair Professorship of Telecommunications at the Georgia Institute of Technology, where he also served as Past Chair of the Telecom Group in ECE and Director of the Broadband Wireless Networking Laboratory. A visionary entrepreneur, he founded Truva Inc., a leading consulting firm based in Georgia, USA, in 1989, and continues to serve as its President. He is a key advisor/consultant to global institutions, including the Technology Innovation Institute (TII) in Abu Dhabi and Odine Labs in Istanbul. Since August 2020, he has been the Founding Editor-in-Chief of the International Telecommunication Union Journal on Future and Evolving Technologies, further solidifying his influence in cutting-edge technology research. Throughout his illustrious career, he has established international research centers in Spain, South Africa, Finland, Saudi Arabia, Russia, and India, fostering innovation across continents. His groundbreaking work now focuses on molecular communication, particularly algae-based networks, alongside pioneering contributions to Networking 2030, holographic and extended-reality communications, 6G/7G wireless systems, terahertz technology, and underwater communications. As of March 2026, his scholarly impact is reflected in an H-index of 146 and more than 155,000 citations according to Google Scholar, underscoring his status as a global leader in telecommunications and networking research. Dr. Akyildiz was elected a Fellow of the ACM in 1997. He has received numerous prestigious awards, including the Humboldt Award (Germany) and the TÜBİTAK Award (Türkiye).
\end{IEEEbiography}

\begin{IEEEbiography}[{\includegraphics[width=1in,height=1.3in,clip,keepaspectratio]{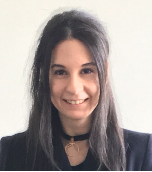}}]{Tuğçe Bilen}, \textit{Member, IEEE},
received her B.Sc., M.Sc., and PhD degrees from the Computer Engineering Department of Istanbul Technical University (ITU), Istanbul, Turkey, in 2015, 2017, and 2022, respectively. During her graduate studies, she served as a Research and Teaching Assistant in the same department. She is currently an Assistant Professor in the Department of Artificial Intelligence and Data Engineering at ITU. Her PhD thesis has been recognised with several prestigious honours, including the IEEE Turkey Section PhD Thesis Award in 2025, the First Prize in Science and Technology from the Turkish Academy of Sciences (TÜBA) in 2023, the Serhat Özyar Young Scientist of the Year Honorary Award in 2023, and the ITU Best PhD Thesis Award in 2022. Her research interests include 6G networks, Knowledge-Defined Networking (KDN), artificial intelligence for network management, and digital twins, with a specific focus on integrating intelligent systems into future network architectures. Dr. Bilen also serves as a reviewer for several reputable international journals.\end{IEEEbiography}

\end{document}